\documentclass[aps,prd,amsfonts,amssymb,twocolumn,nofootinbib]{revtex4}

\usepackage{bm}
\usepackage{graphicx}
\usepackage[small]{caption}
\usepackage{color}

\newcommand{\Tg}{{T_\gamma}}
\newcommand{\ions}[2]{{\text{{\sc #1}\,{\sc #2}}}}
\newcommand{\pot}[2]{#1 \times 10^{#2}}
\newcommand{\Xe}{{X_{\rm e}}}
\newcommand{\fH}{{f_{\rm H}}}
\newcommand{\fHXe}{{f_{\rm H}^{X_{\rm e}}}}
\newcommand{\fHCl}{{f_{\rm H}^{C_{\ell}}}}

\begin{document}

\title{{\sc Rico}: An Accurate Cosmological Recombination Code}

\author{W.~A.~Fendt}
\email{fendt@illinois.edu}
\affiliation{Department of Physics, UIUC, 1110 W.~Green Street, Urbana, IL 61801}

\author{J.~Chluba}
\email{jchluba@mpa-garching.mpg.de}
\affiliation{Max-Planck-Institut f\"{u}r Astrophysik, Karl-Schwarzschild-Str.~1, 85741 
             Garching bei M\"{u}nchen, Germany}

\author{J.~A.~Rubi\~{n}o-Mart\'{i}n}
\email{jalberto@iac.es}
\affiliation{Instituto de Astrof\'{i}sica de Canarias (IAC), C/V\'{i}a L\'{a}ctea, s/n, E-38200, 
             La Laguna, Tenerife (Spain)}

\author{B.~D.~Wandelt}
\email{bwandelt@illinois.edu}
\affiliation{Department of Physics, UIUC, 1110 W.~Green Street, Urbana, IL 61801}
\affiliation{Department of Astronomy, UIUC, 1002 W.~Green Street, Urbana, IL 61801}
\affiliation{Center for Advanced Studies, UIUC, 912 W.~Illinois Street, Urbana, IL 61801}


\begin{abstract}
We present {\sc Rico}, a code designed to compute the ionization 
fraction of the Universe during the epoch of hydrogen and helium recombination
with an unprecedented combination of speed and accuracy.
This is accomplished by training the machine learning code 
{\sc Pico} on the
calculations of a multi-level cosmological recombination code which
self-consistently includes several physical
processes that were neglected previously.
After training, {\sc Rico} is used to fit the free electron fraction as a
function of the cosmological parameters.
While, for example at low redshifts ($z\lesssim 900$), much of the net
change in the ionization fraction can be captured by lowering the hydrogen
fudge factor in {\sc Recfast} by about $3\%$, {\sc Rico}
provides a means of effectively using the accurate ionization history of the
full recombination code in the standard cosmological parameter estimation
framework without the need to add new or refined fudge factors or
functions to a simple recombination model.
Within the new approach presented here it is easy to update {\sc Rico} whenever a 
more accurate full recombination code becomes available. 
Once trained, {\sc Rico} computes the cosmological ionization history with 
negligible fitting error in $\sim10$ milliseconds, a speed-up of 
at least $10^6$ over the full recombination code that was used here. 
Also {\sc Rico} is able to 
reproduce the ionization history of the full code 
to a level well below $0.1\%$, thereby
ensuring that the theoretical power
spectra of CMB fluctuations can be computed to sufficient accuracy and speed for
analysis from upcoming CMB experiments like Planck.
Furthermore it will enable cross-checking different recombination codes
across cosmological parameter space, a comparison that will be very important 
in order to assure the accurate interpretation of future cosmic microwave background data.
\end{abstract}


\maketitle

\section{Introduction}
\label{intro}
Planck, the third generation of satellite missions to study the Cosmic
Microwave Background (CMB), is scheduled for launch later this year.
It will make accurate measurements of the temperature fluctuations in 
the CMB out
to $\ell\sim2500$ as well as provide a detailed picture of the CMB
polarization to $\ell\sim1500$ \citep{Planck2006}.  In order to maximize the
knowledge gained about the underlying cosmological parameters from this high
resolution experiment, it is important that our theoretical calculations are
equally precise.  Currently, uncertainty in the ionization history of the
Universe remains one of the major factors limiting the accuracy of
power spectrum calculations \citep[e.g. see][]{Hu1995, Seljak2003}.

  This fact has recently motived several studies on high precision
  computations of the cosmological hydrogen 
and helium 
recombination history (see Sect.~\ref{sec:Rec_JC} for a detailed overview).
The largest corrections thus far have been due to the non-equilibrium distribution
of 
electrons within the hydrogen sub-states, the absorption of helium photons by neutral
hydrogen and the $2^3 {\rm P}_1-1^1 {\rm S}_0$
triplet-singlet transition in helium.
  Each of the additional physical processes that have been discussed in the
  literature lead to $\gtrsim 0.1\%$ level corrections of the ionization
  history. While the individual changes partially cancel each other,  
  the corresponding overall theoretical uncertainty in the CMB temperature and
  polarization power spectra, in particular at large $\ell$, still exceeds the
  level of $0.1\%$.
  It is also clear that some additional steps must still be taken,
  particularly in connection with the radiative transfer in hydrogen and the
  proper inclusion of two-photon processes (see Sect.~\ref{sec:additions} for
  more details).
These issues are still largely open questions and the subject of intensive ongoing study.

However, the simultaneous and self-consistent inclusion of all these processes
makes the computations very difficult and time-consuming.
It involves the simultaneous evolution of hundreds or thousands of
differential equations for the occupation numbers of the individual levels of
the hydrogen and helium atoms.  As an accurate calculation of the ionization
history of a single cosmological model using the current version
of our full multi-level recombination code takes many hours or even
days of computational time, standard parameter estimation methods require some
approximation of the ionization history that is fast to evaluate.  Currently
the standard method used to evaluate the ionization fraction is the {\sc
Recfast} code \citep{Seager1999}, which
models the hydrogen and helium atoms as effective $3$-level systems.  In order
to approximate the computations done using their multi-level
recombination code \citep{Seager2000}, the authors introduced a fudge factor,
$\fH$, to artificially speed up hydrogen recombination by about $14\%$.  More
recently, \citet{Wong2008} updated {\sc Recfast} to include a second fudge
factor, $b_{\rm He}$, to modify the recombination of helium.  This change,
which allows speeding up \ions{He}{i} recombination, was motivated by
the results of \citet{Switzer2008a}, \citet{Kholupenko2007} and
\citet{Rubino-Mart'in2007}. It is intended to include the effect of
the hydrogen continuum opacity during neutral helium recombination (see
Sect.~\ref{subsec:hopacity}).

Here we provide a different approach to computing a fast and accurate 
approximation of the
recombination history. Our method uses a regression code based on {\sc Pico}
\citep{Fendt2007,Fendt2007a}, which we will henceforth refer to as
{\sc Rico}, to model the ionization history, $X_{\rm e}(z)=N_{\rm
    e}(z)/N_{\rm H}(z)$, as a function of the cosmological parameters.
Here $N_{\rm e}$ denotes the number density of free electrons, and
  $N_{\rm H}$ is the number density of hydrogen nuclei in our Universe.
Since $X_{\rm e}$ is a smooth function, the polynomial approximation
used by {\sc Rico} is extremely accurate, and one retains nearly all of the
information contained in the multi-level code used for training, the {\it training code}.

As the accuracy of the training code is the major determinant of the accuracy
of {\sc Rico}'s approximation, in Sect.~\ref{sec:Rec_JC} we review
the physical processes included in our training code as well as an
assessment of the processes that must still be considered and are
  currently under investigation.
In Sect.~\ref{sec:recfast} we compare the ionization history and corresponding
power spectra computed with the multi-level recombination
  code used for the training of {\sc Rico} to the different versions of {\sc
  Recfast}.  Also we include a short discussion of how one could modify the
approach employed by {\sc Recfast} to capture the new physics included in our
full code.  In Sect.~\ref{sec:Rico} we show that the simple regression scheme
used by {\sc Rico} can accurately model the current full multi-level
recombination calculation, giving power spectra that are sufficiently accurate
for even a cosmic variance limited experiment.

Although the training code discussed here does take into account most of the
important corrections discussed in the literature so far, it does not yet solve
the problem completely. 
Still more physical processes 
should be included and the results validated by independent codes.
As {\sc Rico} can be trained equally well on {\it any} recombination
code it can facilitate crosschecks between those codes as well as a study of how approximations in these codes
propagate to cosmological parameter constraints.  

Our target is to ensure that the ionization history 
can ultimately be calculated to sufficient precision to avoid biasing parameter
estimation from the next generation of CMB experiments. 
By making it easy to propagate advances in the calculation of the ionization
history through to predictions of the CMB power spectra with
{\sc Rico}, 
future development can focus on the physics of recombination and to a lesser degree on the computational efficiency of the training code.

{\sc Rico} and its future updates will be made available at
\verb+http://cosmos.astro.uiuc.edu/rico+.


\section{Description of the multi-level recombination code used for training}
\label{sec:Rec_JC}
\begin{figure}
\centering 
\includegraphics[width=0.99\columnwidth]{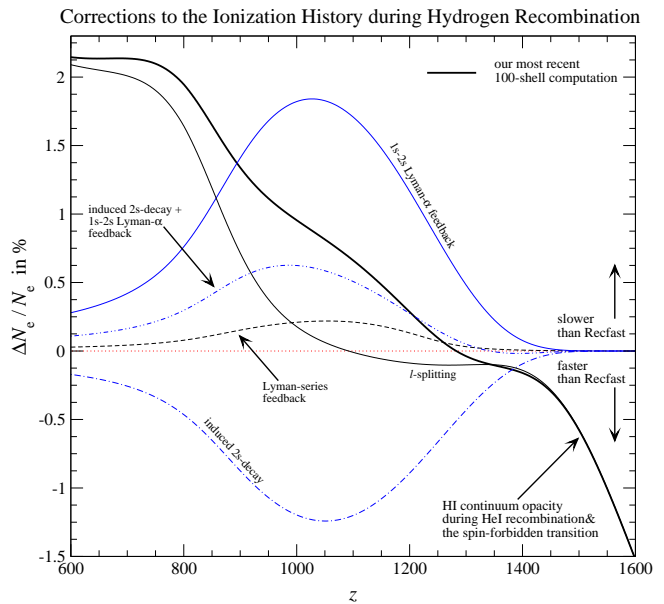}
\caption{
Corrections to the hydrogen recombination history.
For the curves labeled '$l$-splitting' and 'our most recent 100-shell
   computation' we compared with the {\sc Recfast}-code v1.3 after
   removing all of the switches (see Sect. \ref{sec:switches}).
   In the other cases we
   compared with the solution obtained with our code only including the
   non-equilibrium effects. }
\label{fig:HI.Xe}
\end{figure}

In this section we discuss the previously neglected physical processes that
are accounted for in the multi-level recombination code\footnote{This code was
developed as an extension of the works of \citet{Rubino-Mart'in2006} and
\citet{Chluba2007}.} which here is used to train {\sc Rico}.
As we will show, the addition of these processes lead to small but significant
changes in the CMB power spectra (see Sect.~\ref{sec:recfast}).
A comparison of the impact of all the included processes on the recombination
history is illustrated in Fig.~\ref{fig:HI.Xe} and \ref{fig:HeI.Xe}.
In Sect.~\ref{sec:additions} we also give an overview and short discussion of
processes that will be included in a future update of {\sc Rico}.

\subsection{Processes included during Hydrogen recombination}
\label{sec:Hlist}
\subsubsection{Non-equilibrium populations of angular momentum sub-states} 
While the previous work of \citet{Seager2000} evolved the individual
  energy levels of Hydrogen, thereby allowing departures from
Saha-equilibrium, it was assumed that for principal quantum number
  $n>2$ the angular momentum sub-states were in full statistical equilibrium
with each other.
Generally, equilibrium between $l$ sub-states is maintained through collisional
interactions.  During recombination, however, collisions are much weaker than
radiative processes, leading to a departure from full statistical equilibrium
\citep{Rubino-Mart'in2006, Chluba2007}.  

In the computations of \citet{Chluba2007} this lead to a $\sim 5-10\%$
  increase in the free electron fraction, $X_{\rm e}$, at very low redshift
  ($z\lesssim 400$).
  Furthermore, a $\sim 0.6\%$ decrease in $X_{\rm e}$ at decoupling was found in
  comparison with {\sc Recfast}.
  As pointed out in their work this latter correction was {\it not} connected
  with the departures from full statistical equilibrium but has been
  interpreted as a hint towards limitations of the effective 3-level approach
  used in {\sc Recfast} \citep{Chluba2007}.

For hydrogen here we follow the evolution of up to 100 $l$-resolved
shells.
We also include the $l$-changing collisional rates, however their
impact on the recombination history is very small ($\lesssim
0.01\%$ with a maximum at $z\sim 900$).
This problem involves the simultaneous evolution of up to $\sim 5000$
very stiff differential equations;
a task that is numerically very cumbersome
even on modern machines.
The detailed treatment of this process is one of the most difficult and
time-consuming aspects of the whole training code.

Here we have now also updated {\it all} the physical
constants according to the {\sc Nist}-database\footnote{http://www.nist.gov/,
May 2008.}. We then found that the $0.6\%$ decrease in $X_{\rm e}$ at
decoupling practically disappeared. This is mainly because
previously the effect of the reduced mass of hydrogen, which leads to a $\sim
0.05\%$ correction of the ionization potential, was
neglected in the code of \citet{Chluba2007}.  

We have also improved the evaluation of the photoionization and recombination
rates, which for many levels takes rather long. In \citet{Chluba2007}
  the recombination rates for all levels were tabulated before the actual
  computation and {\it detailed balance} was used to infer the photoionization
  rates each time the system is evaluated.  
This treatment is possible as long as the photon and electron temperatures do
not depart significantly from each other, but becomes less accurate at low
redshifts ($z\lesssim 800$).
Here we use the more general procedure described in \citet{Chluba2008b} and
found excellent agreement with the full computation, but at significantly
lower computational cost.

Furthermore, we also discovered a numerical inaccuracy in connection with the
computation of the recombination rates at low redshifts, present in
the code of \citet{Chluba2007}. Fixing this problem decreases the final
correction at $z\lesssim 600-800$ by a factor of $\sim 2$.  However,
this change is not very important in connection with the CMB power spectra,
and also the results for the hydrogen recombination spectrum remain
practically unaltered.

The final result for the changes in the ionization fraction connected
  with the departures from full statistical equilibrium is shown in
  Fig.~\ref{fig:HI.Xe} (curve labeled '$l$-splitting').
To probe the robustness of this result, we carried out two sets of
  comparisons with the program developed by \citet{Rubino-Mart'in2006}.
  Already in the initial version of this code the most recent physical
  constants from the {\sc Nist}-database and also the effect of the
  reduced mass were included.
We explored the robustness of the low-redshift behavior by using
  detailed computations with $n\sim 3$ hydrogen shells down to redshifts
  $z\gtrsim 200$. We found agreement at the level of $\lesssim
  0.001\%$.
As a second set of tests, we focused on the redshift interval around the peak
  of the visibility function, and we compared computations with $n=5$, $10$,
  $15$ and $20$ hydrogen shells. In all cases the agreement
  was better than $0.001\%$ at the redshift range between $z\sim 900$ and $z\sim
  1600$, where the analysis was carried out.
However, for lower redshifts and larger values of $n$ it will still be very
important to have independent confirmation of the result presented here.  

One should also mention that below redshifts $z\sim 200$ we do not follow the
full system anymore, but extend the solution to $z=0$ using a simple 3-level
atom approach similar to {\sc Recfast}.  In order to smoothly connect the
solution we re-scale the derivative of the electron fraction using the
information from the last point of the full computation.
This approximation should have little effect on CMB anisotropy power spectra.
Also the accurate treatment of recombination in this redshift range should
still include the details of primordial chemistry (see
Sect.~\ref{sec:additions}).

\subsubsection{Induced decay of the ${\rm 2s}$-level of Hydrogen}
\label{sec:induced2s}
The ${\rm 2s}\rightarrow{\rm 1s}$ two photon transition plays an important role in the
recombination of hydrogen \citep{Zeldovich1968, Peebles1968} as it provides
one of the primary channels for creating neutral hydrogen\footnote{About
$57\%$ of all hydrogen atoms became neutral via this route, and only $\sim
43\%$ through the Lyman-$\alpha$ line \citep{Chluba2006a}}. This decay rate
is generally computed assuming no background photon field.  However, the
background of CMB photons in the Universe leads to stimulated transitions of
the $2$s-state to the ground level.
Assuming that the CMB is given by a pure blackbody with temperature
$\Tg=2.725\,(1+z)\,$K, it was shown that the induced emission leads to a
$\sim1\%$ increase in the ${\rm 2s}\rightarrow{\rm 1s}$ transition rate
\citep{Chluba2006}.  This increase has the effect of speeding up hydrogen
recombination leading to a maximum change in the free electron fraction of
about $1.3\%$ at $z\sim1050$ (see Fig.~\ref{fig:HI.Xe}, curve labeled 'induced
2s-decay').

Since only the pure CMB blackbody contributes significantly to the
stimulated ${\rm 2s}\rightarrow{\rm 1s}$ transition, it is possible to
tabulate the effective transition rate as a function of temperature before the
computation.
However, for the inverse process (see Sect.~\ref{sec:feed2s}) the
Lyman-$\alpha$ spectral distortion is very important, so we also include it
here for consistency. This is accomplished by first running a 5-shell
computation of the hydrogen recombination to obtain a sufficient estimate of
the Lyman-$\alpha$ distortion within the considered cosmology.
The effective ${\rm 2s}\rightarrow{\rm 1s}$ transition rate is then tabulated
at 5000 redshift points and log-interpolated. The integrals were 
evaluated with relative accuracy $\epsilon\sim 10^{-8}$.

\subsubsection{Feedback of Lyman-$\alpha$ photons on the effective ${\rm 1s}\rightarrow{\rm 2s}$
  absorption rate of Hydrogen}
\label{sec:feed2s}
Due to the large number of photons
produced in Lyman-$\alpha$ transitions during hydrogen recombination there is
a huge excess of radiation over the background spectrum in the far Wien tail
of the CMB.
As noted by
\citet{Kholupenko2006}, 
after some redshifting, these super-Planckian photons are able to re-excite
electrons from the ground-state to the 2s-level, and therefore increase the
effective ${\rm 1s}\rightarrow{\rm 2s}$ absorption rate. This decrease in the
rate of hydrogen recombination (see Fig.~\ref{fig:HI.Xe}, curve labeled '1s-2s
Lyman-$\alpha$ feedback')
practically 
cancels the effect of stimulated 2s-decays (see Sect.~\ref{sec:induced2s}),
leading to a net $\sim 0.6\%$ increase in the free electron fraction at
$z\sim980$ (see Fig.~\ref{fig:HI.Xe}, curve labeled 'induced 2s-decay + 1s-2s
Lyman-$\alpha$ feedback').
While the low redshift behavior of this last result differs slightly
from the curve given in \citet{Kholupenko2006}, it was also recently
obtained by \citet{Hirata2008a}.

As explained in Sect.~\ref{sec:induced2s}, in order to include the
Lyman-$\alpha$ feedback we run a 5-shell computation of hydrogen
recombination to obtain a sufficient estimate of the Lyman-$\alpha$ distortion
and then tabulate the ${\rm 1s}\rightarrow{\rm 2s}$ absorption rate.
We checked that using more shells for the simple run does not affect the
results significantly.

\subsubsection{Feedback within the \ions{H}{i} Lyman-series}
\label{sec:Lyn}
Due to redshifting, all of the Lyman-series photons emitted in the transition of
electrons from levels with $n>2$ have to pass through the next lower-lying
Lyman transition, leading to additional feedback corrections like in the case
of Lyman-$\alpha$ absorption in the 2s-1s two-photon continuum
\citep{Chluba2007a}.
However, in 
this case the photons connected with Ly$n$ are completely absorbed by the
Ly$(n-1)$ resonance and eventually all Ly$n$ photons are converted into
Lyman-$\alpha$ or 2s-1s two-photon decay quanta.
This process delays hydrogen recombination, leading to a maximal correction to
the ionization history of $\Delta N_{\rm e}/N_{\rm e}\sim 0.22\%$ at $z\sim
1050$ (see Fig.~\ref{fig:HI.Xe}, curve labeled 'Lyman-series feedback').
This result was recently confirmed by \citet{Kholupenko2008a}.

To include this feedback
we save the spectral distortion due to
emission of photons in the Lyman-series up to some given $n_{\rm feed}$
(typically we use $n_{\rm feed}\sim 5$) after each time step.
We then include the distortion in the evaluation of the excitation rate of the
next lower-lying Lyman-series transition, assuming that all photons entering
the resonance are absorbed.
Due to the huge optical depth in the Lyman-series this procedure is
well-justified.
In this way the Lyman-series feedback always works like
${\rm Ly}n\,\rightarrow\,{\rm Ly}(n-1)$.
Note that due to the distance between neighboring Lyman-series resonances, the
feedback occurs after the time it takes to redshift into the next transition.

\begin{figure}
\centering 
\includegraphics[width=0.99\columnwidth]{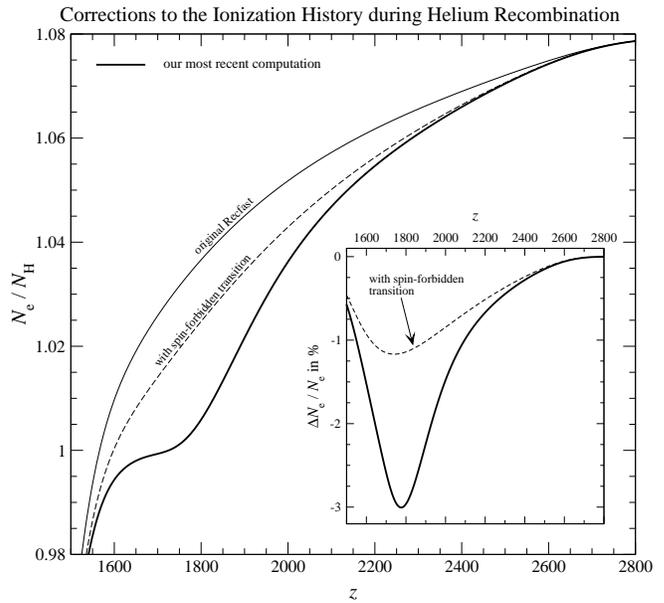}
\caption{
  Corrections to helium recombination history.
  For the inlay plot we compared the ionization history with the
  standard {\sc Recfast}-code v1.3 after removing all of the switches
  (see Sect. \ref{sec:switches}).}
\label{fig:HeI.Xe}
\end{figure}
\subsection{Processes included during Helium recombination}
\label{sec:Helist}
For \ions{He}{ii}-recombination a list of processes similar to hydrogen can be
formulated.
In particular C.~Hirata and E.~Switzer have rigorously
studied these and several additional processes \citep{Switzer2008,
  Hirata2008, Switzer2008a}.
However, since the details of helium recombination history are not strongly
propagating to the computation of the CMB
power spectra, as a first step here we only include the two
dominant corrections, which actually are due to processes that have no analog
in the case of hydrogen recombination. Additional corrections (see discussion
in Sect.~\ref{sec:additions}) will be taken into account for a later release
of {\sc Rico}.
One should also mention that for the computation involving helium we only
include 5 $j$-resolved shells for the singlet and triplet states \citep[for
more detail see][]{Rubino-Mart'in2007}.
We found that within the current numerical approach and related approximations
this leads to sufficient precision (see Sect.~\ref{sec:incompleteness}).

\subsubsection{Spin-forbidden \ions{He}{i}-$2^3 {\rm P}_1-1^1 {\rm S}_0$ transition}
Like for the Lyman-$\alpha$ photons in hydrogen, due to the low expansion rate
of the Universe, \ions{He}{i}-$2^1 {\rm P}_1-1^1 {\rm S}_0$ photons have a
very hard time escaping from the interaction with neutral helium during
$\ions{He}{ii}\rightarrow\ions{He}{i}$ recombination. Therefore
also the recombination of helium is strongly delayed
as compared to the Saha case, and the \ions{He}{i}-$2^1 {\rm
S}_0-1^1 {\rm S}_0$ two-photon transition \citep[$A_{2\gamma}\sim 51.4\,{\rm
s^{-1}}$;][]{Drake1969a} plays a very important role in defining the rate of
\ions{He}{i}-recombination.
Within the standard computation the \ions{He}{i}-$2^1 {\rm S}_0-1^1 {\rm S}_0$
two-photon decay channel allowed $\sim 31\%$ of helium atoms to recombine,
while the \ions{He}{i}-$2^1 {\rm P}_1-1^1 {\rm S}_0$ transition contributed
$\sim 69\%$ \citep[e.g. see table in][]{Wong2007}.

\citet{Dubrovich2005} realized that in addition to the \ions{He}{i}-$2^1 {\rm
  S}_0-1^1 {\rm S}_0$ two-photon transition, the spin-forbidden
  \ions{He}{i}-$2^3 {\rm P}_1-1^1 {\rm S}_0$ transition helps
  helium to recombine even though it also has a very low probability
  \citep[$A\sim 177\,{\rm s^{-1}}$;][]{Drake1969, Lach2001} as
  compared to the \ions{He}{i}-$2^1 {\rm P}_1-1^1 {\rm S}_0$ resonance ($A\sim
  \pot{1.80}{9}\,{\rm s^{-1}}$).
This process, which was not included in the original version of 
{\sc Recfast} \citep{Seager1999}, leads to a $\sim 1.1\%$ decrease in the
free electron fraction at $z\sim 1750$.
Note that in terms of the helium recombination history this corresponds to a
very large effect (see Fig.~\ref{fig:HeI.Xe}).

By including the spin-forbidden \ions{He}{i}-$2^3 {\rm P}_1-1^1 {\rm S}_0$
transitions into the recombination code one finds that about $\sim 17\%$ of
all helium atom became neutral through the \ions{He}{i}-$2^1 {\rm S}_0-1^1
{\rm S}_0$ two-photon transition, $\sim 40\%$ via the \ions{He}{i}-$2^1 {\rm
  P}_1-1^1 {\rm S}_0$ transition and $\sim 43\%$ through this new channel
\citep{Wong2007}.
The comparison shows that the spin-forbidden \ions{He}{i}-$2^3 {\rm P}_1-1^1
{\rm S}_0$ transition is actually one of the dominant channels for helium
recombination.
This conclusion is true even when the hydrogen continuum opacity-effect (see
section \ref{subsec:hopacity}) is also included.  In this case only $8\%$ of
all helium atoms recombine via the \ions{He}{i}-$2^1 {\rm S}_0-1^1
{\rm S}_0$ two-photon decay, while the rest is shared between the
\ions{He}{i}-$2^1 {\rm P}_1-1^1 {\rm S}_0$ and spin-forbidden
\ions{He}{i}-$2^3 {\rm P}_1-1^1 {\rm S}_0$ transition, with a $\sim 2\%$
contribution from other direct transitions to the ground-state
\citep{Rubino-Mart'in2007}.

This process can be self-consistently included by simply adding another term
similar to the normal resonance transitions to the set of rate equations
\citep{Dubrovich2005}, a modification that is now also accounted
for in {\sc Recfast} v1.4 \citep{Wong2008}.
Note that for this transition the Sobolev-escape probability
should be taken into account, since the Sobolev optical depth still reaches
$\sim 3$ close to its maximum.

\subsubsection{Absorption of \ions{He}{i} photons by neutral hydrogen}
\label{subsec:hopacity}
In addition, the recombination of neutral helium is sped up due to the
absorption of \ions{He}{i}-$2^1{\rm P}_1-1^1{\rm S}_0$ and $2^3{\rm
P}_1-1^1{\rm S}_0$-photons by the tiny fraction of neutral hydrogen that is
already present at redshifts $z\lesssim 2400$. 
While photons are emitted close to optically thick \ions{He}{i}-resonances, a
small part of them can be absorbed in the Lyman-continuum of hydrogen,
allowing additional electrons to settle to the ground state of helium.
This process was mentioned by P.~J.~E. Peebles in the mid 90's \citep[see
remark in][]{Hu1995}, but only recently taken into account by
\citet{Switzer2008} and others \citep{Kholupenko2007,
  Rubino-Mart'in2007}.
It leads to the largest correction during the epoch of
$\ions{He}{ii}\rightarrow \ions{He}{i}$-recombination that has been
investigated so far in the literature, strongly accelerating the recombination
of helium below $z\sim 2000$ (see Fig.~\ref{fig:HeI.Xe}, thick solid line).

Since this problem is connected with important details in the radiative
transfer of optically thick \ions{He}{i}-$2^1{\rm P}_1-1^1{\rm S}_0$ and
$2^3{\rm P}_1-1^1{\rm S}_0$, it is a very time-consuming task to solve
self-consistently.
\citet{Switzer2008} studied this problem using a Monte-Carlo approach to
solve the quasi-stationary line-transfer problem. They tabulated the escape
probability for different values of the Sobolev-optical depth and hydrogen
continuum opacity and then interpolated these in the actual recombination
calculation.
Independently, \citet{Kholupenko2007} investigated this problem using a
simplified analytical approach, which slightly underestimated the
acceleration of helium recombination.
They provided simple expressions for the correction to the Sobolev escape
probability of the \ions{He}{i}-$2^1{\rm P}_1-1^1{\rm S}_0$ and $2^3{\rm
P}_1-1^1{\rm S}_0$, which were later also used by \citet{Wong2008} to fudge
the helium recombination history.
\citet{Rubino-Mart'in2007} confirmed the results of \citet{Switzer2008}
using the results of a Fokker-Planck approach to solve the line-transfer
problem \citep{Chlubainpre}.
They also gave a simple 1D-integral which neglects the
redistribution of photons by resonance scattering but still reproduces the
correction to the escape probability rather well.
Very recently \citet{Kholupenko2008b} reconsidered this problem
in more detail analytically and found good agreement with
\citet{Switzer2008} and \citet{Rubino-Mart'in2007}.

In this work we follow the approach of \citet{Rubino-Mart'in2007} and include
the acceleration in the helium recombination history using their 1D-integral
expression \citep[see Eq. (B.3) in][]{Rubino-Mart'in2007} for the correction
to the escape probability.
Here we do not include the redshift dependent fudge-function that
was introduced by \citet{Rubino-Mart'in2007} to account for the additional
acceleration because of partial redistribution.
In the future we plan to solve the full time-dependent problem including the
radiative transfer in the optically thick lines, a task that will also become
necessary for hydrogen recombination.

\subsection{Additional processes that will be considered in the future}
\label{sec:additions}
There are a number of additional processes, both for hydrogen and helium
recombination, that have already been addressed in the literature, but were
not taken into account in the training code. As mentioned above, C.~Hirata and
E.~Switzer have rigorously studied several additional subtle processes
(e.g. feedback; two-photon processes; effect of electron scattering; isotope
shift of $^3{\rm He}$) within the context of helium recombination
\citep[see][]{Switzer2008, Hirata2008, Switzer2008a}.
Also for hydrogen recombination additional processes have been
discussed, however not all of these studies are concluded yet.
Some of these additional processes lead to rather small additional
corrections, but others may still be important.
Below we give a short overview about some of the work that must still be done.
We are planning to make a careful survey in the near future.

\subsubsection{Incompleteness of the atomic models}
\label{sec:incompleteness}
At low redshifts ($z\lesssim 800$) the rate of hydrogen recombination is
strongly controlled by the effective recombination coefficient, which itself
depends on the completeness of the atomic model of hydrogen.
Here we included only up to 100 shells in our computations, but as pointed out
earlier \citep{Chluba2007}, in terms of the ionization history this may
still not be enough, and extension to $\sim 200-300$ shells may be required.
However, for the computation of the CMB power spectra these corrections
probably are not very important, as even with 75 shells rather
converged results seem to be obtained (see Sect.~\ref{sec:Rico}).
Pushing to a larger number of $l$-resolved shells is not trivial, and also
more accurate collisional rates may be required.

As mentioned above, for the computation involving helium we only include 5
$j$-resolved shells.
This approximation seems to be sufficient at the current level of precision
since, unlike the case of hydrogen recombination, the dynamics of helium
recombination are much less controlled by the 
effective recombination
rate, which is strongly connected with the completeness of the atomic model
and can be fudged to some level.
For helium recombination the escape of photons is much more crucial.

\subsubsection{Escape of \ions{H}{i} Lyman-$\alpha$ photons}
\label{sec:Lyaesc}
The escape of photons from the optically thick \ions{H}{i} Lyman-$\alpha$ line
is usually modeled using the Sobolev-approximation.
The validity of this approximation during hydrogen recombination 
has been investigated several times \citep[e.g.][]{Grachev1991, Rybicki1994},
but at the one percent-level a full confirmation is still necessary.
For example, \citet{Grachev2008} recently claimed that the line-recoil effect
leads to a $\sim 1.3\%$ speed-up of hydrogen recombination.
Within the quasi-stationary approach this results seems robust
  \citep{Chlubainpre}.  
Given the importance of this problem to the calculation of the power-spectra,
  independent checks are necessary.
  This is now being investigated in detail \citep{Chlubainpre}.

\subsubsection{Two-photon transitions from higher levels}
\label{sec:2gamma}
Along with the $2s\rightarrow1s$ two-photon transition there are also allowed
two-photon transitions from higher levels to the ground state.
Within the context of hydrogen and helium recombination these transitions were
first studied by \citet{Dubrovich2005}.
They predicted a $\sim5\%$ decrease in the free electron fraction at
$z\sim1200$, however in the computations of the effective two-photon decay
rates of the $n$s and $n$d-levels they only included the first non-resonant
term (i.e. due to the dipole matrix element connecting $n{\rm s}/n{\rm
d}\rightarrow n{\rm p}$) in the infinite sum over intermediate states.

Using rate coefficients for the two-photon decay of the 3s and 3d-levels in
hydrogen as computed by \citet{Cresser1986}, \citet{Wong2007} found that
\citet{Dubrovich2005} overestimated the impact of two-photon transitions on
the ionization history by about one order of magnitude.
However, the calculation of \citet{Cresser1986} was incomplete.
For example they did not include the largest non-resonant term (due to the
dipole matrix element connecting $n{\rm s}/n{\rm d}\rightarrow n{\rm p}$) in
their calculations \citep{Chluba2008a}.
Also physically it is very difficult, if not impossible, to separate the
`pure' two-photon decay rate from the resonant contributions \citep[see
discussions in][]{Chluba2008a, Hirata2008a, Karshenboim2008}, e.g. because of
non-classical interference effects.

Later this problem was reinvestigated in more detail, and a lower limit for the
impact of two-photon decays during hydrogen recombination was derived,
implying that a decrease of slightly more than $0.3-0.5\%$ in the free
electron fraction at $z\sim1150$ can still be expected \citep{Chluba2008a}.
This estimate was obtained by taking into account departures of the full
two-photon line profiles from the Lorentzian shape in the very distant,
optically thin part of the red wing of the Lyman-$\alpha$ line.
According to these computations, the two-photon decays from s-states seem to
slow hydrogen recombination down, while those from d-states speed it up. In
addition it was shown that the slight net acceleration of hydrogen
recombination seems to be dominated by the 3s and 3d contribution
\citep{Chluba2008a}.

However, it was pointed out that the final answer can be only given using a
full radiative transfer computation, which also takes into account the effects
of partial frequency redistribution and the feedback of photons from the blue
side of the Lyman-$\alpha$ resonance \citep[e.g. see][]{Chluba2008a}.
Very recently \citet{Hirata2008a} showed that including these aspects of the
problem, along with the induced 2s-decay and the feedback of Lyman-$\alpha$
photons on the effective ${\rm 1s}\rightarrow{\rm 2s}$ absorption rate,
modifications in the ionization history of the order of $\pm 1.3\%$ can be
expected. Also $2{\rm s}\rightarrow1{\rm s}$ Raman-scattering seems to play
some role.
This would be a very important effect and we are currently investigating it.

Also during helium recombination two-photon transitions and Raman-scattering
are important. 
\citet{Hirata2008} investigated the effect of these, as well as the stimulated
$2^1{\rm S}_0\rightarrow 1^1{\rm S}_0$ two-photon decay, feedback by photons
from the \ions{He}{i}-$2^1{\rm P}_1-1^1{\rm S}_0$ and spin-forbidden
\ions{He}{i}-$2^3 {\rm P}_1-1^1 {\rm S}_0$ transition on the $1^1{\rm
S}_0\rightarrow 2^1{\rm S}_0$ absorption rate.
However, the overall changes 
due to the two-photon transitions and Raman-scattering
seem to be insignificant \citep{Hirata2008}.

\subsubsection{Feedback due to helium lines}
The feedback of high frequency photons released during helium recombination
should also affect the dynamics of hydrogen recombination.
Here it is interesting that most of the high energy photons from
$\ions{He}{iii}\rightarrow\ions{He}{ii}$ will be reprocessed by
neutral helium before they can directly affect hydrogen.
Since the number of high frequency $\gamma(\ions{He}{ii})$ photons is
comparable with the number of helium atoms this should still have a rather
strong effect.
Note that although the ionization history during
$\ions{He}{iii}\rightarrow\ions{He}{ii}$-recombination has a very small impact
on the CMB power-spectra, the exact time-dependence of the
$\gamma(\ions{He}{ii})\rightarrow\ions{He}{i}$ feedback is related to how fast
$\ions{He}{iii}\rightarrow\ions{He}{ii}$-recombination occurred, so that it
may still deserve careful investigation.

Also feedback of the \ions{He}{i} line during helium recombination still has
to be included, and was shown to have a notable delaying effect
\citep{Switzer2008}. However in particular the feedback of
\ions{He}{i}-$2^1{\rm P}_1-1^1{\rm S}_0$ photons on the $2^3 {\rm P}_1-1^1
{\rm S}_0$ transition may in addition require a fully time-dependent
treatment, since both resonances are only separated by $\sim 1\%$ in
frequency, or $\sim 600$ Doppler widths relative to the \ions{He}{i}-$2^1{\rm
  P}_1-1^1{\rm S}_0$ resonance at $z\sim 2500$.

Finally, those photons emitted by neutral helium can directly feedback on
hydrogen, but in order to take this feedback into account more detailed
computations of the helium recombination spectrum are required.
This process could still affect the hydrogen recombination history on a level
exceeding 0.1\%.

\subsubsection{Details of the primordial chemistry at low redshifts} For the
initial computations in connection with {\sc Recfast} \citep{Seager2000}
some aspects of the primordial chemistry were included. These have not yet
been taken into account in our code, but are expected to have only a
small effect, mainly at redshifts $z\lesssim 200$.
However, it is in principle easy to include these, as well as extending the chemical
network using updated rate coefficients \citep[e.g. see][]{Schleicher2008}.

\subsubsection{Other small corrections}
\label{sec:smallcorrs}
Similar to the effect of $^3{\rm He}$ on helium recombination
\citep{Switzer2008a}, also deuterium should affect hydrogen recombination. 
Since due to the isotope shift the deuterium Lyman-$\alpha$ resonance
is on the blue side of the hydrogen Lyman-$\alpha$ line, this should slow down
hydrogen recombination.
However, since the abundance of deuterium is so small, and because at $z\sim
1100$ the shift is only $\sim 12$ Doppler-width of the Lyman-$\alpha$ line,
this process is probably negligible.

For the feedback of the CMB spectral distortion generated during
  recombination on the ${\rm 1s}\rightarrow {\rm 2s}$-absorption rate we only
  took into account the hydrogen Lyman-$\alpha$ distortion. However, the high
  frequency part of the 2s-distortion itself may also have to be included,
  probably leading to another small correction to the ionization history.

\begin{figure}[t]
\centering 
\includegraphics[width=0.99\columnwidth]{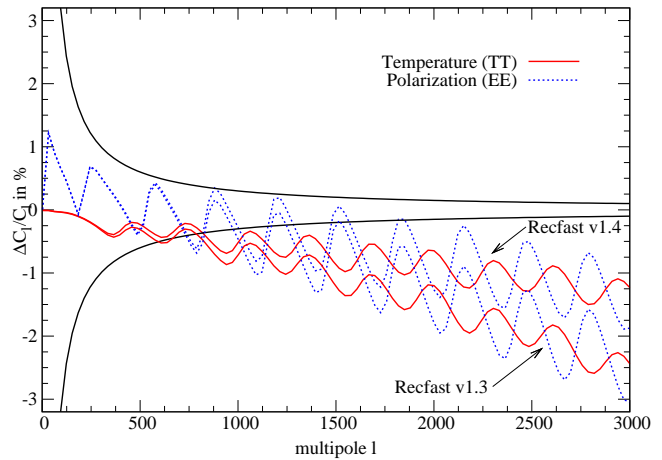}
\caption{Comparison of the angular power spectra computed using the
ionization history of our most recent $100$ shell code versus the
default settings in {\sc Recfast} v1.3 and v1.4.
The two black lines denote our performance benchmark of $\pm 3/\ell$ 
which is roughly $2$ times cosmic variance for the special case of 
fitting a single amplitude parameter (see also \citet{Seljak2003}).
}
\label{fig:ps}
\end{figure}
\section{Comparison with {\sc Recfast}}
\label{sec:recfast}
The net effect of inclusion of the physics listed in
Sects.~\ref{sec:Hlist} and \ref{sec:Helist} leads to an increase in the free
electron fraction during hydrogen recombination. Close to the maximum of the
Thompson visibility function, which is most relevant for computations of the
CMB power spectra, one can observe a $0.7\%$ correction. At low $z$ the error
increases to $\sim 3\%$ at $z\sim 200$.
During helium recombination the correction reaches about $-3\%$ at $z\sim
1800$ (see Fig.~\ref{fig:HeI.Xe} for details).

In Figure \ref{fig:ps} we show the corresponding change in the temperature
and polarization power spectra when comparing with the output of {\sc Recfast}
v1.3 and v1.4 using the default settings.
We used the publicly available code\footnote{http://camb.info} 
CAMB \citep{Lewis2000} to compute the CMB power spectra.
The cosmological model shown has parameter values $\Omega_{\rm B} = 0.0444$,
$\Omega_{\rm M} = 0.2678$, $\Omega_{\rm K} = 0$, $H_0 = 71$ km/s/Mpc, $T_{\rm
cmb} = 2.725$ K, and $Y_{\rm p} = 0.24$.  
The solid and dotted lines show the fractional error between
our $100$ shell calculation and {\sc Recfast} 
for temperature and polarization spectra respectively.
The figure shows that the corrections during helium recombination
contribute about $1/2$ to the average change in the CMB power spectra.
Inclusion of the spin-forbidden \ions{He}{i}-$2^3 {\rm P}_1-1^1 {\rm S}_0$
transitions and fudging of the \ions{He}{i}-$2^1 {\rm P}_1-1^1 {\rm S}_0$
escape probability to account for the effect of the continuum opacity from
neutral hydrogen as done in {\sc Recfast} v1.4 leads to a factor of $2$
improvement in the accuracy of the power spectra.
The remaining error, which is due to inaccuracy in the model of hydrogen used
by {\sc Recfast}, remains above the $0.5\%$ level at $\ell>1000$.  This may
not be sufficient for analysis of Planck data.

Also plotted are dotted lines corresponding to $\pm
3/\ell$, as suggested by \citet{Seljak2003} for a benchmark. These
lines correspond to roughly $2$ times cosmic variance for the special case of
exploring the constraints on a single cosmological parameter.  Since
correlated errors over several $\ell$ values can lead to bias in parameter
constraints even if the error at individual $\ell$'s are less than cosmic
variance, these lines effectively denote an estimate of the minimum error to
which any experiment may be sensitive.  The upcoming Planck satellite mission
should probe the temperature power spectrum beyond $\ell\sim2500$ and be
cosmic variance limited out to $\ell\sim1500$.  Thus the accumulated error in
the power spectrum due to even sub-percent level errors in the ionization
history may have a significant impact on parameter constraints derived from
Planck data, in particular when considering the potential bias on 
inflation parameters (e.g. the spectral index and its running).

\begin{figure}[t]
\centering 
\includegraphics[width=0.99\columnwidth]{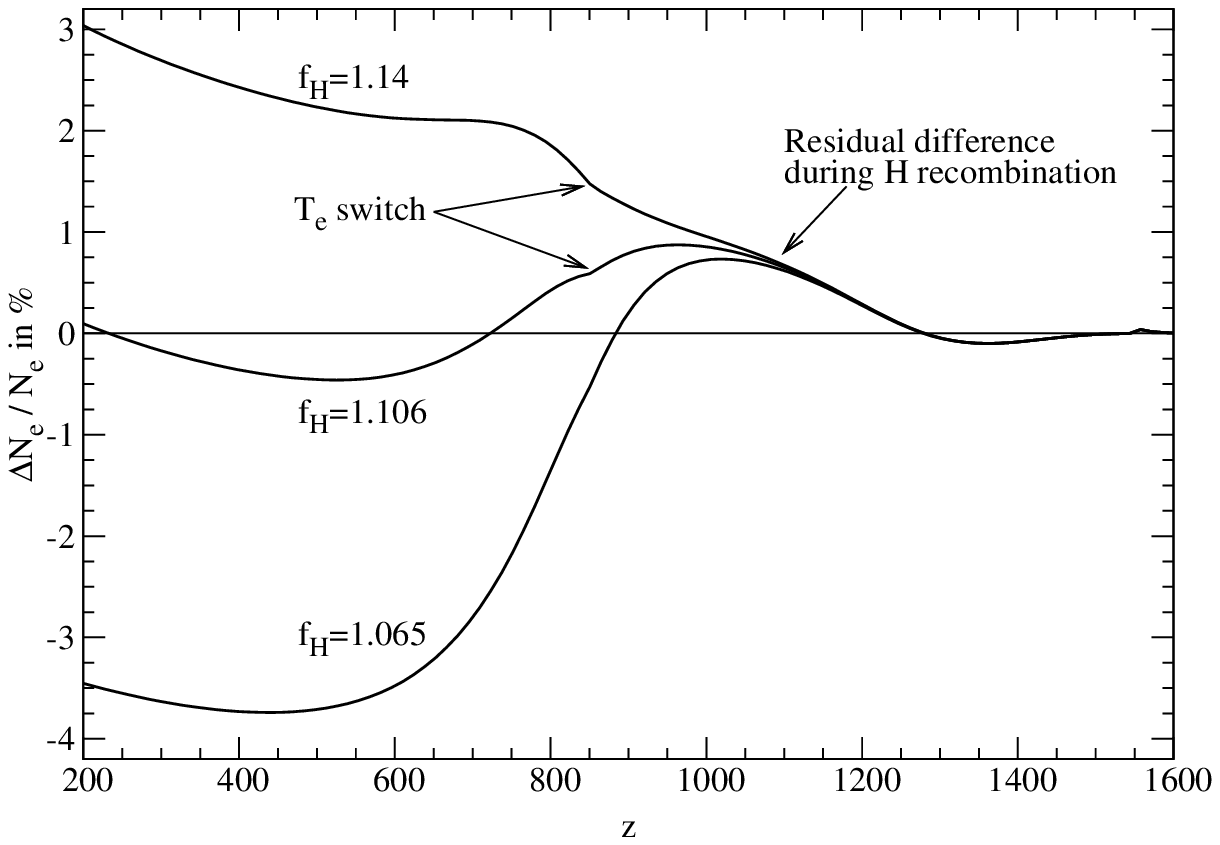}
\\[1mm]
\includegraphics[width=0.99\columnwidth]{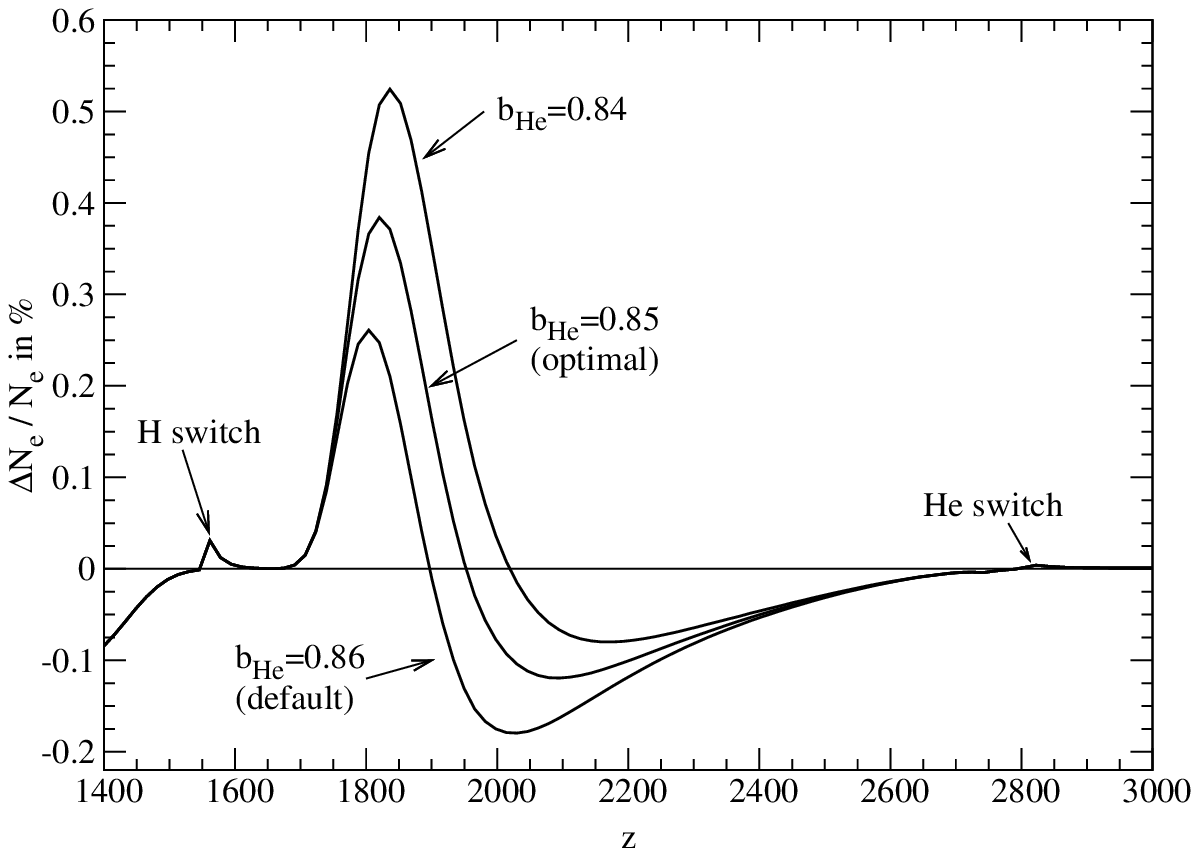}
\caption{Comparison of the ionization fraction during hydrogen (top)
  and helium (bottom) recombination between {\sc Recfast} v1.4 and our
  training code. The ionization history is shown using the default fudge
  factors ($\fH=1.14$, $b_{\rm He}=0.86$) and the value that minimizes the
  fractional error in $X_e$ ($\fHXe=1.106$, $b_{\rm He}=0.85$).
  The top panel also shows the value of $\fH$ that minimizes the cosmic
  variance weighted error in the power spectrum ($\fHCl=1.065$).  
  We have
  also noted the artifacts in the ionization fraction due to the switches
  in {\sc Recfast} (see Sect. \ref{sec:switches}).
}
\label{fig:optfudge}
\end{figure}
\subsection{Modification to the {\sc Recfast} fudge factors}
\label{sec:newfudge}
Some of the improvements to the calculation of the ionization history
discussed in Sects.~\ref{sec:Hlist} and \ref{sec:Helist} can be captured in
{\sc Recfast} by modifying the hydrogen and helium fudge factors, denoted by
$\fH$ and $b_{\rm He}$ respectively. For hydrogen, adjusting $\fH$ to minimize
the fractional error in the ionization history against our most recent $100$
shell calculation over the redshift range $600<z<1200$ gives an optimal value
of $\fHXe=1.106$.
This number is comparable to the one obtained in
  \citet{Rubino-Mart'in2006}, where they found $\fHXe=1.10$ for $30$-shells, and
is roughly a $3\%$ decrease compared to the default value
of $\fH=1.14$ used in {\sc Recfast}. We have performed the same optimization
over $20$ other models and find the optimal value of $\fH$ to lie in the
interval $\left[1.102,1.107\right]$, indicating that there is a
only a small dependence of the optimal fudge factor on the cosmological parameters.  
We note here that, as can be inferred in Fig~\ref{fig:HI.Xe}, the
derived value for $\fH$ is particularly sensitive to the lower limit of the considered
redshift interval.

A similar optimization of the helium fudge factor over redshifts $1400<z<3200$
gives $b_{\rm He}=0.85$, a small change compared to the default value of
$0.86$ used in {\sc Recfast} v1.4.  The modification to the ionization history
from {\sc Recfast} by adjusting these two fudge factors is shown in Figure
\ref{fig:optfudge}.
One should also mention that some differences are expected since the effect
of partial redistribution was neglected here.  This should not have a large
impact but will be considered in a future release of {\sc Rico}.
%

\begin{figure}[t]
\centering 
\includegraphics[width=0.99\columnwidth]{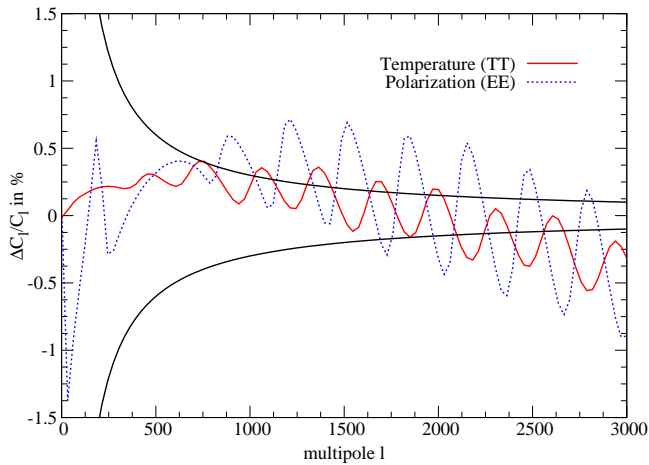}
\caption{Fractional error in the temperature and polarization angular power
  spectra between {\sc Recfast} v1.4 and our training code.  The hydrogen
  fudge factor in {\sc Recfast} is set as $\fHCl=1.065$ which
minimizes the cosmic variance weighted error in the power spectra.
The two solid black lines denote our performance benchmark of
$\Delta C_\ell/C_\ell = 3/\ell$ (see \citet{Seljak2003}). 
}
\label{fig:optcls}
\end{figure}
%
If instead the goal is to minimize the error in the power spectra we find the
optimal value of the hydrogen fudge factor to be $\fHCl=1.065$.  As
shown in Figure \ref{fig:optcls}, using the modified fudge factor captures
some of the changes to the ionization history.  In particular, 
using $\fHCl$ as the fudge factor improves the fit 
in the region where the visibility function is large by sacrificing accuracy
at $z\lesssim800$.  This gives a factor of $2$ improvement in the power spectra 
(see Figure \ref{fig:optcls}).
The further change in helium recombination between our code and {\sc Recfast}
v1.4 has only a tiny effect on the power spectra.

\begin{figure}[t]
\centering 
\includegraphics[width=0.99\columnwidth]{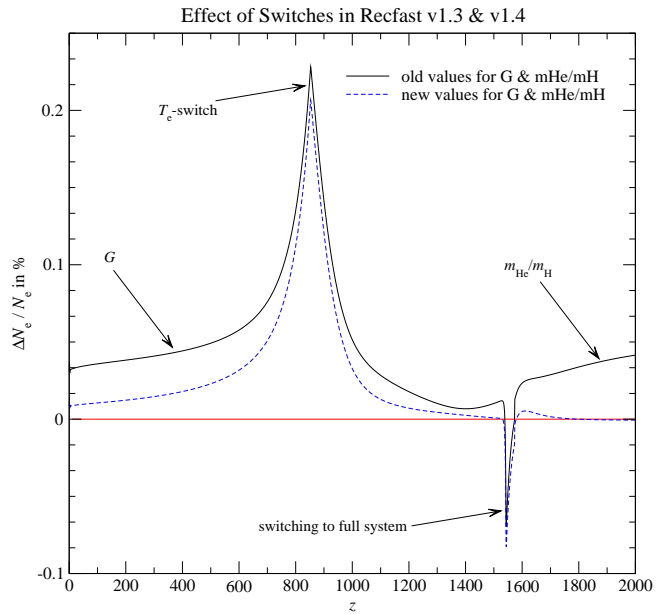}
\caption{The effect of switches in {\sc Recfast} v1.3 and v1.4. For {\sc
    Recfast} v1.4 we have not included any of the recent helium corrections.
  In both cases we compared with our version of {\sc Recfast}, which
  avoids the switches (see Sect.~\ref{sec:switches}). At high redshifts one
  can see the small effect of $m_{\rm He}/m_{\rm H}$.} 
\label{fig:switches}
\end{figure}
\subsection{Effect of the switches in {\sc Recfast}}
\label{sec:switches}
In order to ease the solution of the effective 3-level system used in the {\sc
Recfast} code, some switches were introduced \citep{Seager1999},
that are still present in the current version 1.4.
Starting at redshift $\sim 2800$ for the standard cosmology the {\sc
Recfast}-code only solves the differential equation for helium, using the Saha
solution for hydrogen.
Then when the ratio $N_{\rm p}/N_{\rm H}\lesssim 0.99$ (for the standard
cosmology this occurs at $z\sim 1550$) also the hydrogen equation is solved.
Depending on the derivative for the matter temperature, finally also the
temperature equation is included to the full system (for the standard
cosmology this occurs at $z\sim 820$).

In Fig.~\ref{fig:switches} we show the effect of these switches, which we
avoided in our version\footnote{Note that also the small bug in the
differential equation for hydrogen present in {\sc Recfast} v1.2 was
corrected. This bug leads to a $\sim 0.1\%$ decrease in $\Xe$ at $z\sim
1450$.} of {\sc Recfast}.
For this we changed to a solver for stiff differential equations from
the {\sc Nag}-Library\footnote{See http://www.nag.co.uk/numeric/}.
The most relevant effect is the switch for the temperature equation,
  which is producing a deviation of the order of $\sim 0.2$\% at redshifts
  $z\sim 900$, with respect to our solution in which the temperature equation is
  solved at all redshifts. 
Note that in order to accurately sample the peaks that result from the
switches as well as to maintain consistency with {\sc CAMB} we increased the 
sampling of redshift points used by {\sc Recfast} to $10^{4}$.

For completeness, Fig.~\ref{fig:switches} also presents the
  comparison with the case in which the corrections due to the new values of
  $G$ and $m_{\rm He}/m_{\rm H}$, pointed out by \citet{Wong2007}, are not
  included. These two modifications were introduced in {\sc Recfast} v1.3. One
  can see that at $z\lesssim 2000$ these lead to corrections that are below
  $\sim 0.05\%$, and actually only reach $\sim 0.1\%$ at $z\gtrsim 6000$. 
This modification was already taken into account in the initial
version of the work by \citet{Rubino-Mart'in2006}.

Note that for the comparisons in this section, our ``{\sc Recfast}
computation'' does not take into account any of the helium corrections, which
are incorporated to {\sc Recfast} v1.4 using fudging.

\begin{figure}[t]
\centering 
\includegraphics[width=0.99\columnwidth]{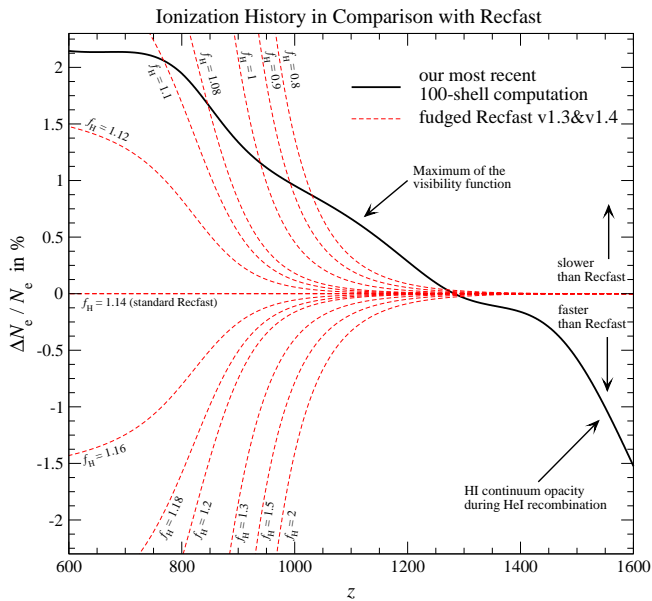}
\caption{Comparison of the ionization history of our most recent $100$-shell
  code with {\sc Recfast} v1.3 after removing all the switches (see
  Sect.~\ref{sec:switches}). Also included are the variations in {\sc Recfast}
  v1.3\&v1.4 for different values of the hydrogen fudge factor.  }
\label{fig:fudge}
\end{figure}

\begin{figure}[t]
\centering 
\includegraphics[width=0.99\columnwidth]{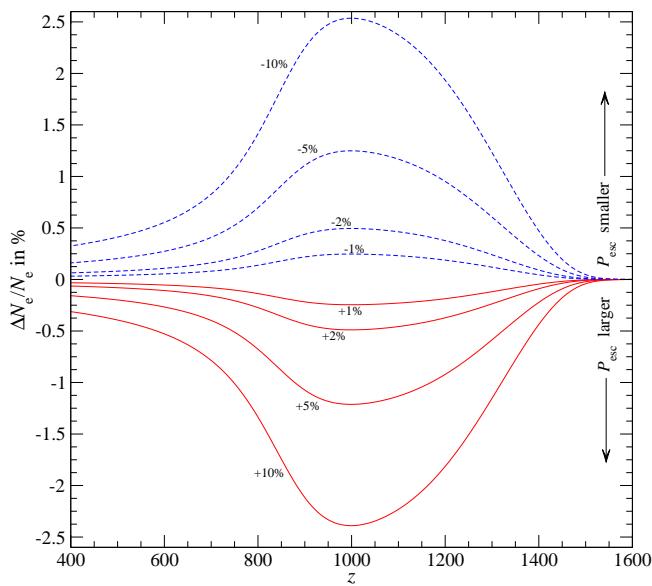}
\caption{Relative change in the number density of free electrons when allowing
for a constant relative change in the Sobolev-escape probability of the
\ions{H}{i} Lyman-$\alpha$ transition as labeled. All curves where computed
using {\sc Recfast}. At maximum $\Delta N_{\rm e}/N_{\rm e}\sim
-\frac{1}{4}\Delta P/P$.}
\label{fig:Pesc}
\end{figure}
\subsection{To what extent does fudging work?}
\label{sec:fudging}
In this Section we discuss our perspective of how far one can go with
fudging, and what kind of modification may help to further improve the result
with a simplified approach similar to {\sc Recfast}.
However, to us `fudging' at some level either becomes equivalent with fitting
the whole problem, or the performance of the corresponding algorithm would
probably also be insufficient for cosmological parameter
estimations.
This is the reason why we decide to take a direct fitting approach (see
Sect.~\ref{sec:Rico}) leading to {\sc Rico}, instead of a detailed
investigation of the possibilities for more sophisticated fudging.

\subsubsection{Fudging hydrogen recombination}
\label{sec:fudgingH}
Physically the fudge factor $\fH$ introduced to speed up hydrogen
recombination increases the effective recombination coefficient to the second
shell.
This implies that the hydrogen fudge factor can only alter the recombination
history significantly when the recombination dynamics truly depend on this
coefficient.
However, as Fig.~\ref{fig:fudge} illustrates this is only the case at
low redshifts.
While to some extent the differences with our full recombination calculation
can be reduced to optimizing the value of $\fH$ (see
Sect.~\ref{sec:newfudge}), clearly not all the differences can be
erased (see Fig.~\ref{fig:optfudge} and \ref{fig:optcls}). In addition there
is some slight dependence of $\fH$ on cosmology (see
Sect.~\ref{sec:newfudge}), which one may still be interested in.

As proposed by \citet{Chluba2007}, the next level of fudging is
using a redshift dependent function for $\fH$. Looking at Fig.~\ref{fig:fudge}
one can realize that in particular at $z\sim 1100$ this fudge-function should
depend rather strongly on redshift, and that for the high redshift part this
approach will likely not lead to very good results.
Additionally the cosmology dependence of this fudge-function will probably be
significant. In the end it is equivalent to directly fitting the correction
for the standard concordance model only.

In order to capture some of the corrections from a more physical point of
view, one could think about simple extensions of the effective 3-level atom.
The induced two-photon decay (see Sect.~\ref{sec:induced2s}) can be
incorporated rather easily, since only the effective decay rate due to
blackbody CMB photons should be replaced \citep{Chluba2006}.
For the feedback of the Lyman-$\alpha$ distortion on the effective 1s-2s
absorption rate (see Sect.~\ref{sec:feed2s}) one should provide an
approximation for the Lyman-$\alpha$ distortion, which in principle can be
done analytically \citep{Kholupenko2006}.
Including these two corrections should therefore be possible.

In order to include the Lyman-series feedback (see Sect.~\ref{sec:Lyn}) one
could simply take more shells, say $5-10$, into account, and hence model this
process more or less self-consistently. However, runs with 5 shells will take
longer,
and one should provide effective recombination coefficients for all the
additional shells in order to correctly model the Lyman-series lines.
An alternative analytic treatment was recently presented by
\citet{Kholupenko2008b}.

In order to include changes close to decoupling one could also modify the
\ions{H}{i} Lyman-$\alpha$ escape rate, or alternatively the
effective two-photon decay rate of the the 2s level. As an example, in
Fig.~\ref{fig:Pesc} we illustrate how changes in the \ions{H}{i}
Lyman-$\alpha$ escape affect the ionization history.
This could certainly help to take into account some more details of the
radiative transfer problem, which are still under discussion and not included
here (see Sect.~\ref{sec:Lyaesc}).
Also the effect of two-photon processes from high levels
(see Sect.~\ref{sec:2gamma}) and to some extent the details of
radiative transfer (see Sect.~\ref{sec:Lyaesc}) could probably be
approximated within such an approach.
However, this will certainly require a carefully calibrated redshift dependent
fudge-function, as was also used by \citet{Rubino-Mart'in2007} in the case of
helium recombination.
Again one then should check the cosmology dependence of this fudge-function,
which, depending on the desired accuracy, may vary significantly.
It is clear that within such an approach the connection to the
real physical processes will no longer be obvious.

\subsubsection{Fudging helium recombination}
\label{sec:fudgingHe}
For helium recombination introducing a fudge-factor for the effective helium
recombination coefficient would not help at all. Even if one increases
the recombination coefficient by more than a factor of two, the helium recombination
history basically does not change.
Therefore one directly has to fudge the escape rate of \ions{He}{i}-$2^1{\rm
P}_1-1^1{\rm S}_0$ and $2^3 {\rm P}_1-1^1 {\rm S}_0$ photons, as it has been
done recently by \citet{Wong2008}.
Again, depending on the desired accuracy, one should allow for a redshift
dependent fudge-function as was already proposed by
\citet{Rubino-Mart'in2007}, since only the low redshift tail of helium
recombination is accurately reproduced by {\sc Recfast} v1.4.
However, since details in the helium recombination history are not strongly
propagating to the computation of the temperature and polarization power
spectra, fudging is probably sufficient from this point of view.
Still, using {\sc Rico} would enable one to reproduce the helium computations
at a much higher level of accuracy and with significantly less effort,
even when more detailed computations become available.

\begin{figure}[t]
\begin{center}
\includegraphics[width=0.99\columnwidth]{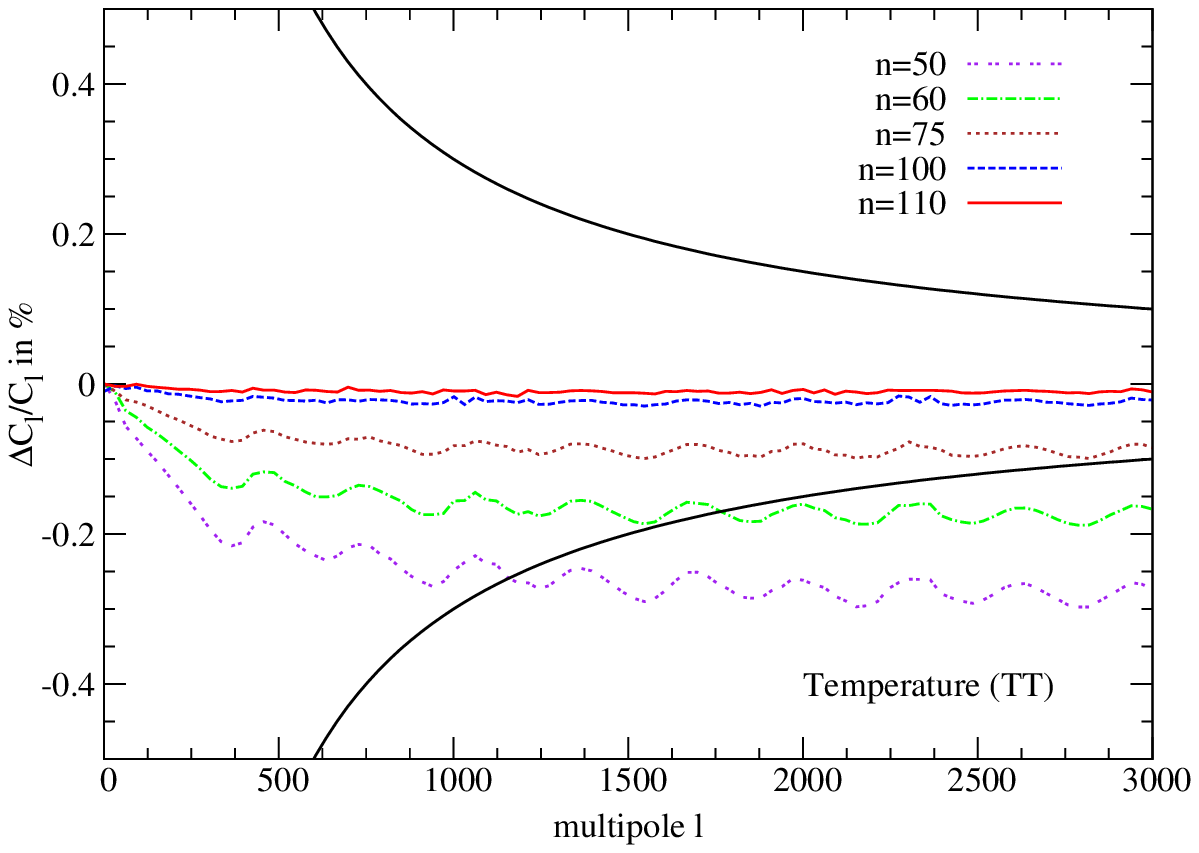}
\includegraphics[width=0.99\columnwidth]{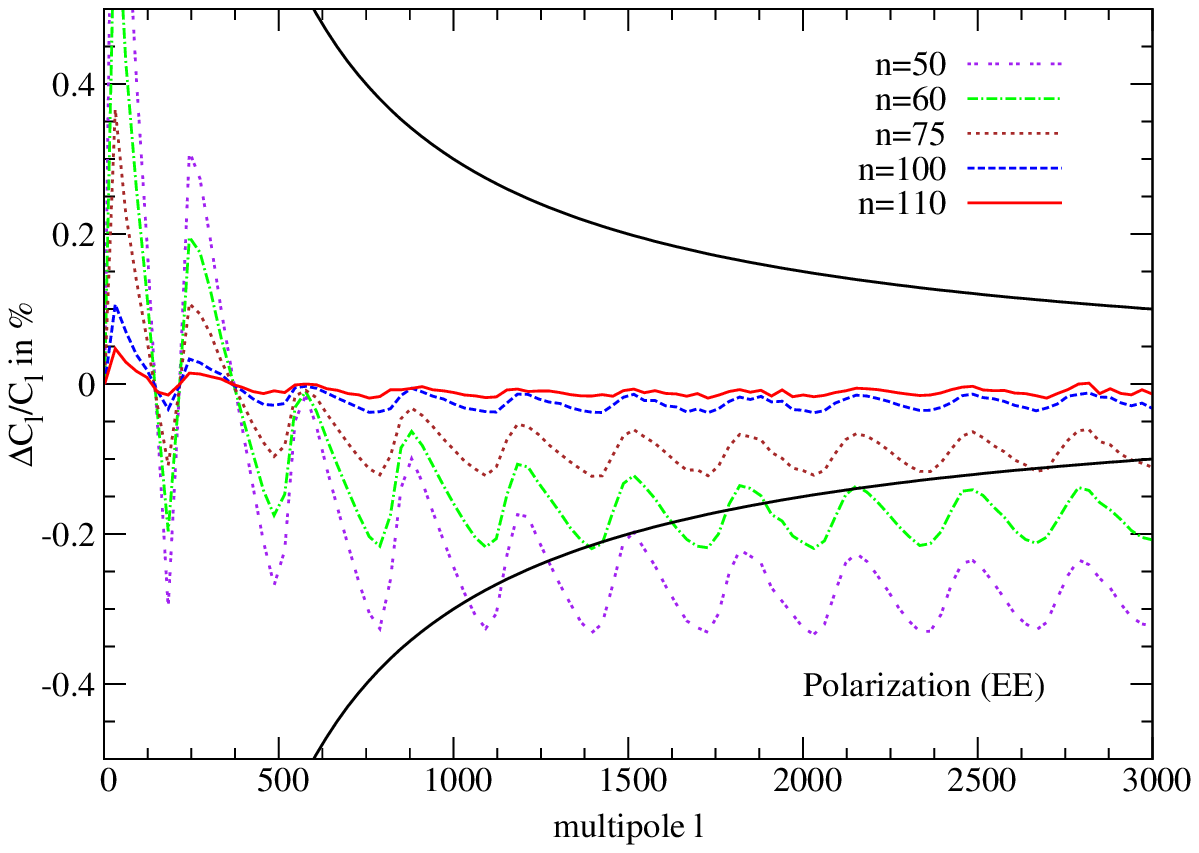}
   \caption{Comparison of the ionization history computed by our code for varying
            numbers of shells.  The fractional error is taken against the $120$ shell
            calculation.  Note that little physics is lost by using the $100$ shell
            approximation compared to the $120$ shell and that the $60$ shell 
            computation remains below our minimum cosmic variance estimate until
            around $\ell\sim2000$.
            The two solid black lines denote our performance benchmark of
            $\Delta C_\ell/C_\ell = 3/\ell$ (see also \citet{Seljak2003}).
            \label{fig:level_comp}}
\end{center}
\end{figure}

\section{Approximating the Full Recombination Calculation}
\label{sec:Rico}
Instead of introducing new fudge factors or modifying existing ones to
reproduce the results of the full recombination code using the simplified
$3$-level model, we propose directly fitting to the results of the accurate
calculation using a training set of ionization histories with a regression
code similar to {\sc Pico} \citep{Fendt2007,Fendt2007a}.  Since the
ionization history is a smooth function of the cosmological parameters, {\sc
Rico} is able to accurately capture most of the variation in this
function.

Ignoring massive neutrinos, the ionization history depends on $6$ cosmological
parameters\footnote{If necessary one could probably reduce the
number of independent variables by changing to a parameterization that encodes
the variables that directly impact the recombination calculation. However
currently this is not a limiting factor.}: the baryon density $\Omega_{\rm B}
h^2$, the dark matter density $\Omega_{\rm CDM} h^2$, the curvature density
$\Omega_{\rm K}$, the Hubble constant $H_0$, the temperature of the CMB
$T_{\rm cmb}$ and the helium mass fraction $Y_{\rm p}$.  The training set for
{\sc Rico} is generated based on constraints from the WMAP 3 year results
\citep{Spergel2007}.  It is therefore convenient to use $\theta$, the ratio of
the sound horizon to the angular diameter distance at decoupling, instead of
$H$ as there will be significantly less correlation in the parameters
\citep{Kosowsky2002}.
The parameters
$\Omega_{\rm B} h^2$, $\Omega_{\rm CDM} h^2$, $\Omega_{\rm K}$ and $\theta$ are
chosen to lie within $25$ log-likelihoods of the peak of the WMAP $3$ year
likelihood.  A method for efficiently finding these points is discussed in
\citet{Fendt2007a}.  Values for the current temperature of the CMB and the
helium fraction are chosen uniformly from the intervals $\left[2.72,2.73\right]$
\citep{Fixsen2002} and $\left[0.232,0.258\right]$ \citep{Olive2004}
respectively.
In the parameter estimation framework, outside of this region we just use 
the standard {\sc Recfast} code. However, these points will have low likelihood
and should not effect the results the results of parameter analysis.

We train {\sc Rico} using $500$ parameter sets and corresponding ionization
histories.  For this work, we chose to use $60$ shell models as they remain
very accurate out to $\ell\sim2000$.  This is shown in Figure
\ref{fig:level_comp} where we have plotted the fractional error using
various numbers of shells compared to the $120$ model.  The two
black lines again correspond to $\pm 3/\ell$.  Note the small difference
between the $100$ and $120$ shell model, indicating that there
is little additional information gained about the power spectra from tracking
shell populations beyond $100$.  Also, the $60$ shell line remains below our
strict error tolerance until roughly $\ell \sim 2000$ indicating that Planck
will not be sensitive to the error from using the lower shell calculations.

\begin{figure}[t]
\begin{center}
\includegraphics[width=0.99\columnwidth]{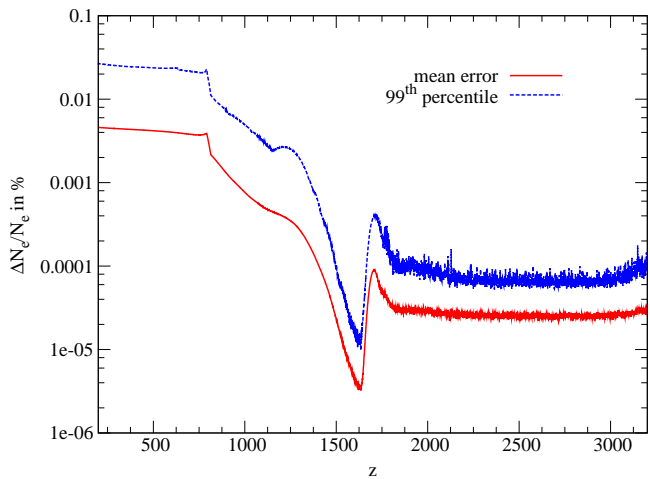}
   \caption{Performance of {\sc Rico} in computing the ionization history based on $500$ 
            training cases computed with $60$ shells.  This test was done over $400$
            separate test cases.  The lines denote the mean fractional error along 
            with the fractional error that bounds $99\%$ of the test cases.
             \label{fig:pico_xe}}
\end{center}
\end{figure}
\begin{figure}[t]
   \begin{center}
   \includegraphics[width=0.99\columnwidth]{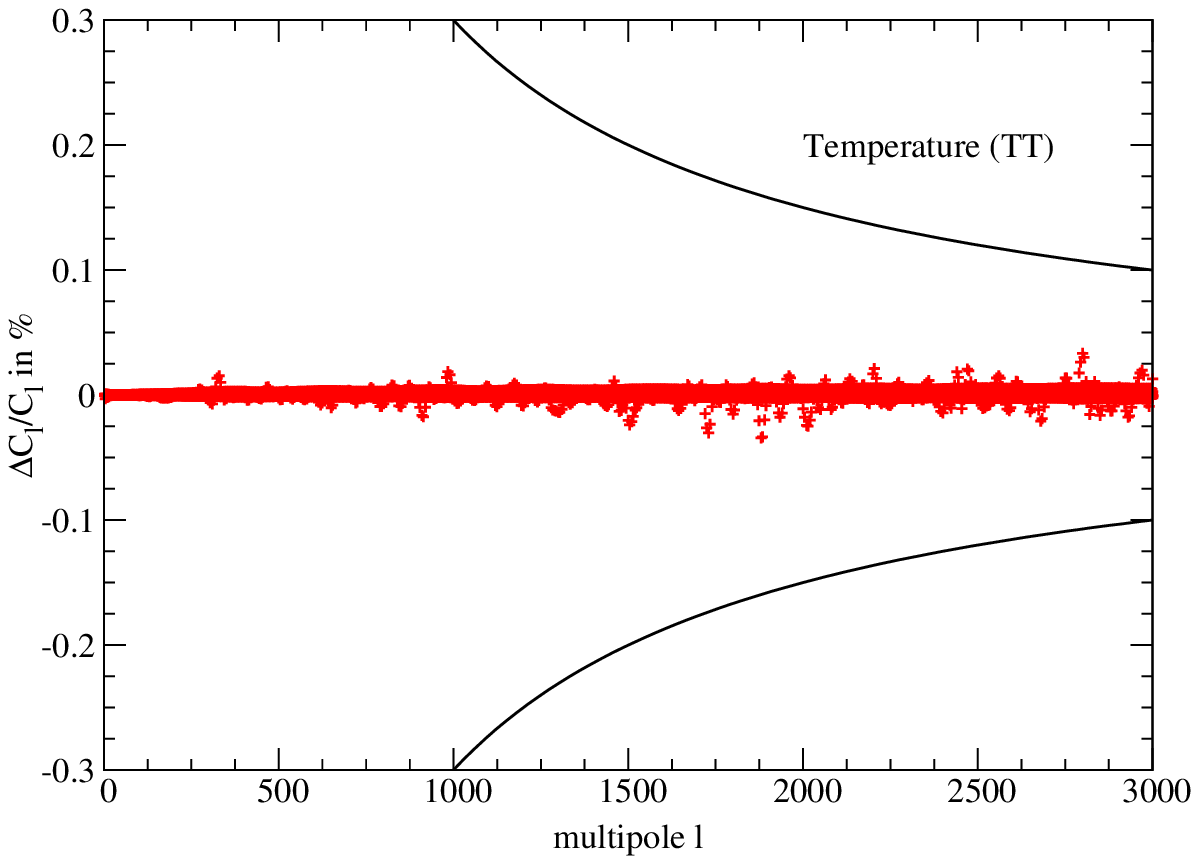}
   \includegraphics[width=0.99\columnwidth]{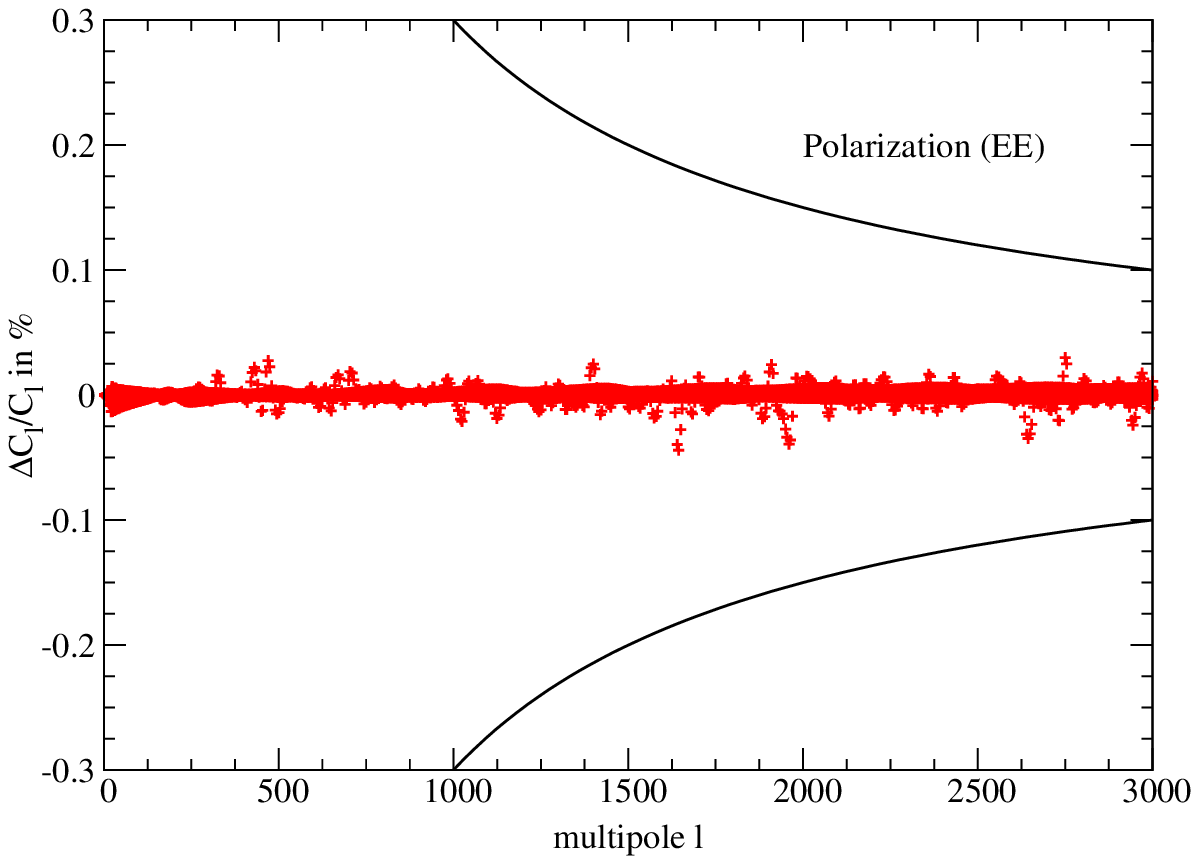}
   \caption{Fractional error in the power spectra using the recombination history
            computed by {\sc Rico}. The plot includes $200$ models from our test 
            set and the comparison has been made with the power spectra using the 
            recombination history from the $60$ shell run of our full recombination
            code.
            The two solid black lines denote our performance benchmark of
            $\Delta C_\ell/C_\ell = 3/\ell$ (see \citet{Seljak2003}).
            \label{fig:pico_cl_err}}
   \end{center}
\end{figure}

%
To test the accuracy of {\sc Rico} we generated a test set of $400$ models
chosen in the same manner as the training set.  The models used for this test
are not included in the training set.  As in the training set, the
recombination histories that make up the test set were run using $60$ shells
in order to access only the error incurred by approximating the ionization
history with {\sc Rico}.  Figure \ref{fig:pico_xe} shows the error in the
ionization history computed by {\sc Rico} against the full recombination code.
At high $z$, there is little variation in the training set so {\sc Rico} has
no trouble approximating the ionization fraction.  After $z\sim1700$, as
\ions{He}{i} recombination begins, until the end of hydrogen recombination the
error from {\sc Rico} remains below $0.03\%$ for $99\%$ of the test cases.
The corresponding error in the power spectra from using {\sc Rico} is thus
negligible (see Fig.~\ref{fig:pico_cl_err}).  

This demonstrates that {\sc Rico} essentially propagates all of the
information from the ionization history into the power spectra. The ability
to accurately compute the $C_\ell$'s is primarily limited by the accuracy of
the full recombination code used to train {\sc Rico}. Since {\sc Rico} uses the
same fitting methodology as {\sc Pico} it is possible to extend the parameter space
or add additional model parameters (for example parameterizing uncertainties in the
recombination calculation). This would increase the one-time training cost, but would 
{\it not}
lead to an appreciable additional cost in the evaluation of 
the ionization history.
If necessary, the accuracy of {\sc Rico} can be further increased by
adding more points to the training set as well as by using a higher order
regression. Again, the only computational penalty is to the one-time 
training step.

While the downside of this method is the requirement of running the full code
$\sim500$ times to generate a training set, this can be done completely in
parallel.  Large computing clusters or distributed computing projects are
ideal for exactly this type of application.  Also, this training cost does not affect
the user of the code. The advantage is that there is no
need to find optimal approximations using a simple physical system. {\sc Rico}
can just be trained using results from the most accurate code available.
Also, {\sc Rico} is trained to compute the ionization history over a volume of
parameter space and not simply optimized based on a single model at the peak
of the parameter likelihood.

\section{Conclusion}
We have presented a new code designed to compute the ionization history of the
Universe.  The code includes previously neglected physics that leads to changes
in the ionization fraction at the $2-3\%$ level in some redshift
regions.  This change leads to a correction to the CMB power spectra of 
more than
$1\%$ at $\ell\sim3000$.  As uncertainty in the ionization fraction is the main
contributor to error in the theoretical CMB anisotropy power spectra, this code represents a
significant step in our ability to compute the power spectra to a significant
precision for upcoming experiments.

While it is possible to attempt to capture the changes to the ionization
fraction by modification of the hydrogen fudge factor in {\sc Recfast}, some
residual error remains near
the peak of the visibility function.  
The problem of correctly introducing
fudge factors can be avoided entirely by using a regression code based on a
training set of cosmological parameters and their corresponding ionization
histories.  A simple polynomial fit is sufficient to compute ionization
fraction at the level of $\sim0.01\%$ over the volume of parameter
space relevant to current experimental data.

As it is straightforward to train {\sc Rico} on multiple recombination codes,
our algorithm enables the propagation of approximations made to the ionization
history to biases on the cosmological parameters when analyzing data.  Since
implementing the many new physical processes already discussed in the
literature in an accurate and robust manner is very challenging, we hope that
this ability facilitates a model by model cross validation among the
recombination codes that have been developed.  
Given the complexity of the physical processes involved in the
computation of the cosmological recombination history, such a comparison will
become very important in order to ensure that the final result is correct.
Our goal is to ensure that the ionization history 
can ultimately be calculated to sufficient precision to avoid biasing parameter
estimation from the next generation of CMB experiments. 
By making it easy to propagate advances in the calculation of the ionization
history through to predictions of the CMB power spectra with
{\sc Rico}, 
future development can focus on the physics of recombination and to a lesser 
degree on the computational efficiency of the physical recombination code.

{\sc Rico} and its future updates will be made available at
\verb+http://cosmos.astro.uiuc.edu/rico+.

\acknowledgments
J.C. wishes to thank R.~A.~Sunyaev for many useful
  discussions and suggestions. 
J.C. is also grateful to D.~Scott for discussion.  The authors
are also grateful to M.~Bucher, V.~Dubrovich, C.~Hirata, 
S.~Karshenboim, E.~Kholupenko, R.~Porter, 
E.~Switzer and W.~Wong for discussion during the Workshop `The Physics of
Cosmological Recombination' at the Max-Planck-Institut f\"{u}r Astrophysik in
Garching, Germany, July 2008.
BDW and WAF were partially supported through NSF award 0507676. BDW was
partially supported by a Friedrich Wilhelm Bessel Prize by the Alexander von
Humboldt Foundation.  WAF was partially supported through a fellowship from
the Computational Science and Engineering Department at the University of
Illinois at Urbana-Champaign.  This work made use of the Department of Physics
computing cluster at the University of Illinois at Urbana-Champaign.

\bibliography{rico}

\begin{thebibliography}{43}
\expandafter\ifx\csname natexlab\endcsname\relax\def\natexlab#1{#1}\fi
\expandafter\ifx\csname bibnamefont\endcsname\relax
  \def\bibnamefont#1{#1}\fi
\expandafter\ifx\csname bibfnamefont\endcsname\relax
  \def\bibfnamefont#1{#1}\fi
\expandafter\ifx\csname citenamefont\endcsname\relax
  \def\citenamefont#1{#1}\fi
\expandafter\ifx\csname url\endcsname\relax
  \def\url#1{\texttt{#1}}\fi
\expandafter\ifx\csname urlprefix\endcsname\relax\def\urlprefix{URL }\fi
\providecommand{\bibinfo}[2]{#2}
\providecommand{\eprint}[2][]{\url{#2}}

\bibitem[{\citenamefont{{The Planck Collaboration}}(2006)}]{Planck2006}
\bibinfo{author}{\bibnamefont{{The Planck Collaboration}}},
  \bibinfo{journal}{ArXiv Astrophysics e-prints}  (\bibinfo{year}{2006}),
  \eprint{astro-ph/0604069}.

\bibitem[{\citenamefont{{Hu} et~al.}(1995)\citenamefont{{Hu}, {Scott},
  {Sugiyama}, and {White}}}]{Hu1995}
\bibinfo{author}{\bibfnamefont{W.}~\bibnamefont{{Hu}}},
  \bibinfo{author}{\bibfnamefont{D.}~\bibnamefont{{Scott}}},
  \bibinfo{author}{\bibfnamefont{N.}~\bibnamefont{{Sugiyama}}},
  \bibnamefont{and} \bibinfo{author}{\bibfnamefont{M.}~\bibnamefont{{White}}},
  \bibinfo{journal}{\prd} \textbf{\bibinfo{volume}{52}}, \bibinfo{pages}{5498}
  (\bibinfo{year}{1995}), \eprint{arXiv:astro-ph/9505043}.

\bibitem[{\citenamefont{{Seljak} et~al.}(2003)\citenamefont{{Seljak},
  {Sugiyama}, {White}, and {Zaldarriaga}}}]{Seljak2003}
\bibinfo{author}{\bibfnamefont{U.}~\bibnamefont{{Seljak}}},
  \bibinfo{author}{\bibfnamefont{N.}~\bibnamefont{{Sugiyama}}},
  \bibinfo{author}{\bibfnamefont{M.}~\bibnamefont{{White}}}, \bibnamefont{and}
  \bibinfo{author}{\bibfnamefont{M.}~\bibnamefont{{Zaldarriaga}}},
  \bibinfo{journal}{\prd} \textbf{\bibinfo{volume}{68}},
  \bibinfo{pages}{083507} (\bibinfo{year}{2003}),
  \eprint{arXiv:astro-ph/0306052}.

\bibitem[{\citenamefont{{Seager} et~al.}(1999)\citenamefont{{Seager},
  {Sasselov}, and {Scott}}}]{Seager1999}
\bibinfo{author}{\bibfnamefont{S.}~\bibnamefont{{Seager}}},
  \bibinfo{author}{\bibfnamefont{D.~D.} \bibnamefont{{Sasselov}}},
  \bibnamefont{and} \bibinfo{author}{\bibfnamefont{D.}~\bibnamefont{{Scott}}},
  \bibinfo{journal}{\apjl} \textbf{\bibinfo{volume}{523}}, \bibinfo{pages}{L1}
  (\bibinfo{year}{1999}), \eprint{arXiv:astro-ph/9909275}.

\bibitem[{\citenamefont{{Seager} et~al.}(2000)\citenamefont{{Seager},
  {Sasselov}, and {Scott}}}]{Seager2000}
\bibinfo{author}{\bibfnamefont{S.}~\bibnamefont{{Seager}}},
  \bibinfo{author}{\bibfnamefont{D.~D.} \bibnamefont{{Sasselov}}},
  \bibnamefont{and} \bibinfo{author}{\bibfnamefont{D.}~\bibnamefont{{Scott}}},
  \bibinfo{journal}{\apjs} \textbf{\bibinfo{volume}{128}}, \bibinfo{pages}{407}
  (\bibinfo{year}{2000}), \eprint{arXiv:astro-ph/9912182}.

\bibitem[{\citenamefont{{Wong} et~al.}(2008)\citenamefont{{Wong}, {Moss}, and
  {Scott}}}]{Wong2008}
\bibinfo{author}{\bibfnamefont{W.~Y.} \bibnamefont{{Wong}}},
  \bibinfo{author}{\bibfnamefont{A.}~\bibnamefont{{Moss}}}, \bibnamefont{and}
  \bibinfo{author}{\bibfnamefont{D.}~\bibnamefont{{Scott}}},
  \bibinfo{journal}{\mnras} \textbf{\bibinfo{volume}{386}},
  \bibinfo{pages}{1023} (\bibinfo{year}{2008}), \eprint{arXiv:0711.1357}.

\bibitem[{\citenamefont{{Switzer} and
  {Hirata}}(2008{\natexlab{a}})}]{Switzer2008a}
\bibinfo{author}{\bibfnamefont{E.~R.} \bibnamefont{{Switzer}}}
  \bibnamefont{and} \bibinfo{author}{\bibfnamefont{C.~M.}
  \bibnamefont{{Hirata}}}, \bibinfo{journal}{\prd}
  \textbf{\bibinfo{volume}{77}}, \bibinfo{pages}{083008}
  (\bibinfo{year}{2008}{\natexlab{a}}), \eprint{arXiv:astro-ph/0702145}.

\bibitem[{\citenamefont{{Kholupenko} et~al.}(2007)\citenamefont{{Kholupenko},
  {Ivanchik}, and {Varshalovich}}}]{Kholupenko2007}
\bibinfo{author}{\bibfnamefont{E.~E.} \bibnamefont{{Kholupenko}}},
  \bibinfo{author}{\bibfnamefont{A.~V.} \bibnamefont{{Ivanchik}}},
  \bibnamefont{and} \bibinfo{author}{\bibfnamefont{D.~A.}
  \bibnamefont{{Varshalovich}}}, \bibinfo{journal}{\mnras}
  \textbf{\bibinfo{volume}{378}}, \bibinfo{pages}{L39} (\bibinfo{year}{2007}),
  \eprint{arXiv:astro-ph/0703438}.

\bibitem[{\citenamefont{{Rubi{\~n}o-Mart{\'{\i}}n}
  et~al.}(2007)\citenamefont{{Rubi{\~n}o-Mart{\'{\i}}n}, {Chluba}, and
  {Sunyaev}}}]{Rubino-Mart'in2007}
\bibinfo{author}{\bibfnamefont{J.~A.}
  \bibnamefont{{Rubi{\~n}o-Mart{\'{\i}}n}}},
  \bibinfo{author}{\bibfnamefont{J.}~\bibnamefont{{Chluba}}}, \bibnamefont{and}
  \bibinfo{author}{\bibfnamefont{R.~A.} \bibnamefont{{Sunyaev}}},
  \bibinfo{journal}{ArXiv e-prints} \textbf{\bibinfo{volume}{711}}
  (\bibinfo{year}{2007}), \eprint{0711.0594}.

\bibitem[{\citenamefont{{Fendt} and {Wandelt}}(2007{\natexlab{a}})}]{Fendt2007}
\bibinfo{author}{\bibfnamefont{W.~A.} \bibnamefont{{Fendt}}} \bibnamefont{and}
  \bibinfo{author}{\bibfnamefont{B.~D.} \bibnamefont{{Wandelt}}},
  \bibinfo{journal}{\apj} \textbf{\bibinfo{volume}{654}}, \bibinfo{pages}{2}
  (\bibinfo{year}{2007}{\natexlab{a}}), \eprint{arXiv:astro-ph/0606709}.

\bibitem[{\citenamefont{{Fendt} and
  {Wandelt}}(2007{\natexlab{b}})}]{Fendt2007a}
\bibinfo{author}{\bibfnamefont{W.~A.} \bibnamefont{{Fendt}}} \bibnamefont{and}
  \bibinfo{author}{\bibfnamefont{B.~D.} \bibnamefont{{Wandelt}}},
  \bibinfo{journal}{ArXiv e-prints}  (\bibinfo{year}{2007}{\natexlab{b}}),
  \eprint{0712.0194}.

\bibitem[{\citenamefont{{Rubi{\~n}o-Mart{\'{\i}}n}
  et~al.}(2006)\citenamefont{{Rubi{\~n}o-Mart{\'{\i}}n}, {Chluba}, and
  {Sunyaev}}}]{Rubino-Mart'in2006}
\bibinfo{author}{\bibfnamefont{J.~A.}
  \bibnamefont{{Rubi{\~n}o-Mart{\'{\i}}n}}},
  \bibinfo{author}{\bibfnamefont{J.}~\bibnamefont{{Chluba}}}, \bibnamefont{and}
  \bibinfo{author}{\bibfnamefont{R.~A.} \bibnamefont{{Sunyaev}}},
  \bibinfo{journal}{\mnras} \textbf{\bibinfo{volume}{371}},
  \bibinfo{pages}{1939} (\bibinfo{year}{2006}),
  \eprint{arXiv:astro-ph/0607373}.

\bibitem[{\citenamefont{{Chluba} et~al.}(2007)\citenamefont{{Chluba},
  {Rubi{\~n}o-Mart{\'{\i}}n}, and {Sunyaev}}}]{Chluba2007}
\bibinfo{author}{\bibfnamefont{J.}~\bibnamefont{{Chluba}}},
  \bibinfo{author}{\bibfnamefont{J.~A.}
  \bibnamefont{{Rubi{\~n}o-Mart{\'{\i}}n}}}, \bibnamefont{and}
  \bibinfo{author}{\bibfnamefont{R.~A.} \bibnamefont{{Sunyaev}}},
  \bibinfo{journal}{\mnras} \textbf{\bibinfo{volume}{374}},
  \bibinfo{pages}{1310} (\bibinfo{year}{2007}),
  \eprint{arXiv:astro-ph/0608242}.

\bibitem[{\citenamefont{{Chluba} and
  {Sunyaev}}(2008{\natexlab{a}})}]{Chluba2008b}
\bibinfo{author}{\bibfnamefont{J.}~\bibnamefont{{Chluba}}} \bibnamefont{and}
  \bibinfo{author}{\bibfnamefont{R.~A.} \bibnamefont{{Sunyaev}}},
  \bibinfo{journal}{ArXiv e-prints} \textbf{\bibinfo{volume}{803}}
  (\bibinfo{year}{2008}{\natexlab{a}}), \eprint{0803.3584}.

\bibitem[{\citenamefont{{Zeldovich} et~al.}(1968)\citenamefont{{Zeldovich},
  {Kurt}, and {Syunyaev}}}]{Zeldovich1968}
\bibinfo{author}{\bibfnamefont{Y.~B.} \bibnamefont{{Zeldovich}}},
  \bibinfo{author}{\bibfnamefont{V.~G.} \bibnamefont{{Kurt}}},
  \bibnamefont{and} \bibinfo{author}{\bibfnamefont{R.~A.}
  \bibnamefont{{Syunyaev}}}, \bibinfo{journal}{Zhurnal Eksperimental noi i
  Teoreticheskoi Fiziki} \textbf{\bibinfo{volume}{55}}, \bibinfo{pages}{278}
  (\bibinfo{year}{1968}).

\bibitem[{\citenamefont{{Peebles}}(1968)}]{Peebles1968}
\bibinfo{author}{\bibfnamefont{P.~J.~E.} \bibnamefont{{Peebles}}},
  \bibinfo{journal}{\apj} \textbf{\bibinfo{volume}{153}}, \bibinfo{pages}{1}
  (\bibinfo{year}{1968}).

\bibitem[{\citenamefont{{Chluba} and
  {Sunyaev}}(2006{\natexlab{a}})}]{Chluba2006a}
\bibinfo{author}{\bibfnamefont{J.}~\bibnamefont{{Chluba}}} \bibnamefont{and}
  \bibinfo{author}{\bibfnamefont{R.~A.} \bibnamefont{{Sunyaev}}},
  \bibinfo{journal}{\aap} \textbf{\bibinfo{volume}{458}}, \bibinfo{pages}{L29}
  (\bibinfo{year}{2006}{\natexlab{a}}), \eprint{arXiv:astro-ph/0608120}.

\bibitem[{\citenamefont{{Chluba} and
  {Sunyaev}}(2006{\natexlab{b}})}]{Chluba2006}
\bibinfo{author}{\bibfnamefont{J.}~\bibnamefont{{Chluba}}} \bibnamefont{and}
  \bibinfo{author}{\bibfnamefont{R.~A.} \bibnamefont{{Sunyaev}}},
  \bibinfo{journal}{\aap} \textbf{\bibinfo{volume}{446}}, \bibinfo{pages}{39}
  (\bibinfo{year}{2006}{\natexlab{b}}), \eprint{arXiv:astro-ph/0508144}.

\bibitem[{\citenamefont{{Kholupenko} and {Ivanchik}}(2006)}]{Kholupenko2006}
\bibinfo{author}{\bibfnamefont{E.~E.} \bibnamefont{{Kholupenko}}}
  \bibnamefont{and} \bibinfo{author}{\bibfnamefont{A.~V.}
  \bibnamefont{{Ivanchik}}}, \bibinfo{journal}{Astronomy Letters}
  \textbf{\bibinfo{volume}{32}}, \bibinfo{pages}{795} (\bibinfo{year}{2006}),
  \eprint{arXiv:astro-ph/0611395}.

\bibitem[{\citenamefont{{Hirata}}(2008)}]{Hirata2008a}
\bibinfo{author}{\bibfnamefont{C.~M.} \bibnamefont{{Hirata}}},
  \bibinfo{journal}{ArXiv e-prints} \textbf{\bibinfo{volume}{803}}
  (\bibinfo{year}{2008}), \eprint{0803.0808}.

\bibitem[{\citenamefont{{Chluba} and {Sunyaev}}(2007)}]{Chluba2007a}
\bibinfo{author}{\bibfnamefont{J.}~\bibnamefont{{Chluba}}} \bibnamefont{and}
  \bibinfo{author}{\bibfnamefont{R.~A.} \bibnamefont{{Sunyaev}}},
  \bibinfo{journal}{\aap} \textbf{\bibinfo{volume}{475}}, \bibinfo{pages}{109}
  (\bibinfo{year}{2007}), \eprint{arXiv:astro-ph/0702531}.

\bibitem[{\citenamefont{{Kholupenko}
  et~al.}(2008{\natexlab{a}})\citenamefont{{Kholupenko}, {Ivanchik}, and
  {Varshalovich}}}]{Kholupenko2008a}
\bibinfo{author}{\bibfnamefont{E.~E.} \bibnamefont{{Kholupenko}}},
  \bibinfo{author}{\bibfnamefont{A.~V.} \bibnamefont{{Ivanchik}}},
  \bibnamefont{and} \bibinfo{author}{\bibfnamefont{D.~A.}
  \bibnamefont{{Varshalovich}}}, \bibinfo{journal}{Astronomy Letters
  (accepted)}  (\bibinfo{year}{2008}{\natexlab{a}}).

\bibitem[{\citenamefont{{Switzer} and
  {Hirata}}(2008{\natexlab{b}})}]{Switzer2008}
\bibinfo{author}{\bibfnamefont{E.~R.} \bibnamefont{{Switzer}}}
  \bibnamefont{and} \bibinfo{author}{\bibfnamefont{C.~M.}
  \bibnamefont{{Hirata}}}, \bibinfo{journal}{\prd}
  \textbf{\bibinfo{volume}{77}}, \bibinfo{pages}{083006}
  (\bibinfo{year}{2008}{\natexlab{b}}), \eprint{arXiv:astro-ph/0702143}.

\bibitem[{\citenamefont{{Hirata} and {Switzer}}(2008)}]{Hirata2008}
\bibinfo{author}{\bibfnamefont{C.~M.} \bibnamefont{{Hirata}}} \bibnamefont{and}
  \bibinfo{author}{\bibfnamefont{E.~R.} \bibnamefont{{Switzer}}},
  \bibinfo{journal}{\prd} \textbf{\bibinfo{volume}{77}},
  \bibinfo{pages}{083007} (\bibinfo{year}{2008}),
  \eprint{arXiv:astro-ph/0702144}.

\bibitem[{\citenamefont{{Drake} et~al.}(1969)\citenamefont{{Drake}, {Victor},
  and {Dalgarno}}}]{Drake1969a}
\bibinfo{author}{\bibfnamefont{G.~W.~F.} \bibnamefont{{Drake}}},
  \bibinfo{author}{\bibfnamefont{G.~A.} \bibnamefont{{Victor}}},
  \bibnamefont{and}
  \bibinfo{author}{\bibfnamefont{A.}~\bibnamefont{{Dalgarno}}},
  \bibinfo{journal}{Physical Review} \textbf{\bibinfo{volume}{180}},
  \bibinfo{pages}{25} (\bibinfo{year}{1969}).

\bibitem[{\citenamefont{{Wong} and {Scott}}(2007)}]{Wong2007}
\bibinfo{author}{\bibfnamefont{W.~Y.} \bibnamefont{{Wong}}} \bibnamefont{and}
  \bibinfo{author}{\bibfnamefont{D.}~\bibnamefont{{Scott}}},
  \bibinfo{journal}{\mnras} \textbf{\bibinfo{volume}{375}},
  \bibinfo{pages}{1441} (\bibinfo{year}{2007}),
  \eprint{arXiv:astro-ph/0610691}.

\bibitem[{\citenamefont{{Dubrovich} and {Grachev}}(2005)}]{Dubrovich2005}
\bibinfo{author}{\bibfnamefont{V.~K.} \bibnamefont{{Dubrovich}}}
  \bibnamefont{and} \bibinfo{author}{\bibfnamefont{S.~I.}
  \bibnamefont{{Grachev}}}, \bibinfo{journal}{Astronomy Letters}
  \textbf{\bibinfo{volume}{31}}, \bibinfo{pages}{359} (\bibinfo{year}{2005}).

\bibitem[{\citenamefont{{Drake} and {Dalgarno}}(1969)}]{Drake1969}
\bibinfo{author}{\bibfnamefont{G.~W.~F.} \bibnamefont{{Drake}}}
  \bibnamefont{and}
  \bibinfo{author}{\bibfnamefont{A.}~\bibnamefont{{Dalgarno}}},
  \bibinfo{journal}{\apj} \textbf{\bibinfo{volume}{157}}, \bibinfo{pages}{459}
  (\bibinfo{year}{1969}).

\bibitem[{\citenamefont{{{\L}ach} and {Pachucki}}(2001)}]{Lach2001}
\bibinfo{author}{\bibfnamefont{G.}~\bibnamefont{{{\L}ach}}} \bibnamefont{and}
  \bibinfo{author}{\bibfnamefont{K.}~\bibnamefont{{Pachucki}}},
  \bibinfo{journal}{\pra} \textbf{\bibinfo{volume}{64}},
  \bibinfo{pages}{042510} (\bibinfo{year}{2001}),
  \eprint{arXiv:physics/0105110}.

\bibitem[{\citenamefont{{Chluba} and
  {Sunyaev}}(2008{\natexlab{b}})}]{Chlubainpre}
\bibinfo{author}{\bibfnamefont{J.}~\bibnamefont{{Chluba}}} \bibnamefont{and}
  \bibinfo{author}{\bibfnamefont{R.~A.} \bibnamefont{{Sunyaev}}},
  \bibinfo{journal}{in preparation}  (\bibinfo{year}{2008}{\natexlab{b}}).

\bibitem[{\citenamefont{{Kholupenko}
  et~al.}(2008{\natexlab{b}})\citenamefont{{Kholupenko}, {Ivanchik}, and
  {Varshalovich}}}]{Kholupenko2008b}
\bibinfo{author}{\bibfnamefont{E.~E.} \bibnamefont{{Kholupenko}}},
  \bibinfo{author}{\bibfnamefont{A.~V.} \bibnamefont{{Ivanchik}}},
  \bibnamefont{and} \bibinfo{author}{\bibfnamefont{D.~A.}
  \bibnamefont{{Varshalovich}}}, \bibinfo{journal}{Communication fr the
  Workshop `The Physics of Cosmological Recombination' Garching}
  (\bibinfo{year}{2008}{\natexlab{b}}).

\bibitem[{\citenamefont{{Grachev} and {Dubrovich}}(1991)}]{Grachev1991}
\bibinfo{author}{\bibfnamefont{S.~I.} \bibnamefont{{Grachev}}}
  \bibnamefont{and} \bibinfo{author}{\bibfnamefont{V.~K.}
  \bibnamefont{{Dubrovich}}}, \bibinfo{journal}{Astrophysics}
  \textbf{\bibinfo{volume}{34}}, \bibinfo{pages}{124} (\bibinfo{year}{1991}).

\bibitem[{\citenamefont{{Rybicki} and {dell'Antonio}}(1994)}]{Rybicki1994}
\bibinfo{author}{\bibfnamefont{G.~B.} \bibnamefont{{Rybicki}}}
  \bibnamefont{and} \bibinfo{author}{\bibfnamefont{I.~P.}
  \bibnamefont{{dell'Antonio}}}, \bibinfo{journal}{\apj}
  \textbf{\bibinfo{volume}{427}}, \bibinfo{pages}{603} (\bibinfo{year}{1994}),
  \eprint{arXiv:astro-ph/9312006}.

\bibitem[{\citenamefont{{Grachev} and {Dubrovich}}(2008)}]{Grachev2008}
\bibinfo{author}{\bibfnamefont{S.~I.} \bibnamefont{{Grachev}}}
  \bibnamefont{and} \bibinfo{author}{\bibfnamefont{V.~K.}
  \bibnamefont{{Dubrovich}}}, \bibinfo{journal}{ArXiv e-prints}
  \textbf{\bibinfo{volume}{801}} (\bibinfo{year}{2008}), \eprint{0801.3347}.

\bibitem[{\citenamefont{Cresser et~al.}(1986)\citenamefont{Cresser, Tang,
  Salamo, and Chan}}]{Cresser1986}
\bibinfo{author}{\bibfnamefont{J.~D.} \bibnamefont{Cresser}},
  \bibinfo{author}{\bibfnamefont{A.~Z.} \bibnamefont{Tang}},
  \bibinfo{author}{\bibfnamefont{G.~J.} \bibnamefont{Salamo}},
  \bibnamefont{and} \bibinfo{author}{\bibfnamefont{F.~T.} \bibnamefont{Chan}},
  \bibinfo{journal}{Phys. Rev. A} \textbf{\bibinfo{volume}{33}},
  \bibinfo{pages}{1677} (\bibinfo{year}{1986}).

\bibitem[{\citenamefont{{Chluba} and
  {Sunyaev}}(2008{\natexlab{c}})}]{Chluba2008a}
\bibinfo{author}{\bibfnamefont{J.}~\bibnamefont{{Chluba}}} \bibnamefont{and}
  \bibinfo{author}{\bibfnamefont{R.~A.} \bibnamefont{{Sunyaev}}},
  \bibinfo{journal}{\aap} \textbf{\bibinfo{volume}{480}}, \bibinfo{pages}{629}
  (\bibinfo{year}{2008}{\natexlab{c}}), \eprint{arXiv:0705.3033}.

\bibitem[{\citenamefont{{Karshenboim} and {Ivanov}}(2008)}]{Karshenboim2008}
\bibinfo{author}{\bibfnamefont{S.~G.} \bibnamefont{{Karshenboim}}}
  \bibnamefont{and} \bibinfo{author}{\bibfnamefont{V.~G.}
  \bibnamefont{{Ivanov}}}, \bibinfo{journal}{Astronomy Letters}
  \textbf{\bibinfo{volume}{34}}, \bibinfo{pages}{289} (\bibinfo{year}{2008}).

\bibitem[{\citenamefont{{Schleicher} et~al.}(2008)\citenamefont{{Schleicher},
  {Galli}, {Palla}, {Camenzind}, {Klessen}, {Bartelmann}, and
  {Glover}}}]{Schleicher2008}
\bibinfo{author}{\bibfnamefont{D.~R.~G.} \bibnamefont{{Schleicher}}},
  \bibinfo{author}{\bibfnamefont{D.}~\bibnamefont{{Galli}}},
  \bibinfo{author}{\bibfnamefont{F.}~\bibnamefont{{Palla}}},
  \bibinfo{author}{\bibfnamefont{M.}~\bibnamefont{{Camenzind}}},
  \bibinfo{author}{\bibfnamefont{R.~S.} \bibnamefont{{Klessen}}},
  \bibinfo{author}{\bibfnamefont{M.}~\bibnamefont{{Bartelmann}}},
  \bibnamefont{and} \bibinfo{author}{\bibfnamefont{S.~C.~O.}
  \bibnamefont{{Glover}}}, \bibinfo{journal}{ArXiv e-prints}
  \textbf{\bibinfo{volume}{803}} (\bibinfo{year}{2008}), \eprint{0803.3987}.

\bibitem[{\citenamefont{{Lewis} et~al.}(2000)\citenamefont{{Lewis},
  {Challinor}, and {Lasenby}}}]{Lewis2000}
\bibinfo{author}{\bibfnamefont{A.}~\bibnamefont{{Lewis}}},
  \bibinfo{author}{\bibfnamefont{A.}~\bibnamefont{{Challinor}}},
  \bibnamefont{and}
  \bibinfo{author}{\bibfnamefont{A.}~\bibnamefont{{Lasenby}}},
  \bibinfo{journal}{\apj} \textbf{\bibinfo{volume}{538}}, \bibinfo{pages}{473}
  (\bibinfo{year}{2000}), \eprint{arXiv:astro-ph/9911177}.

\bibitem[{\citenamefont{{Spergel} et~al.}(2007)\citenamefont{{Spergel}, {Bean},
  {Dor{\'e}}, {Nolta}, {Bennett}, {Dunkley}, {Hinshaw}, {Jarosik}, {Komatsu},
  {Page} et~al.}}]{Spergel2007}
\bibinfo{author}{\bibfnamefont{D.~N.} \bibnamefont{{Spergel}}},
  \bibinfo{author}{\bibfnamefont{R.}~\bibnamefont{{Bean}}},
  \bibinfo{author}{\bibfnamefont{O.}~\bibnamefont{{Dor{\'e}}}},
  \bibinfo{author}{\bibfnamefont{M.~R.} \bibnamefont{{Nolta}}},
  \bibinfo{author}{\bibfnamefont{C.~L.} \bibnamefont{{Bennett}}},
  \bibinfo{author}{\bibfnamefont{J.}~\bibnamefont{{Dunkley}}},
  \bibinfo{author}{\bibfnamefont{G.}~\bibnamefont{{Hinshaw}}},
  \bibinfo{author}{\bibfnamefont{N.}~\bibnamefont{{Jarosik}}},
  \bibinfo{author}{\bibfnamefont{E.}~\bibnamefont{{Komatsu}}},
  \bibinfo{author}{\bibfnamefont{L.}~\bibnamefont{{Page}}},
  \bibnamefont{et~al.}, \bibinfo{journal}{\apjs}
  \textbf{\bibinfo{volume}{170}}, \bibinfo{pages}{377} (\bibinfo{year}{2007}),
  \eprint{arXiv:astro-ph/0603449}.

\bibitem[{\citenamefont{{Kosowsky} et~al.}(2002)\citenamefont{{Kosowsky},
  {Milosavljevic}, and {Jimenez}}}]{Kosowsky2002}
\bibinfo{author}{\bibfnamefont{A.}~\bibnamefont{{Kosowsky}}},
  \bibinfo{author}{\bibfnamefont{M.}~\bibnamefont{{Milosavljevic}}},
  \bibnamefont{and}
  \bibinfo{author}{\bibfnamefont{R.}~\bibnamefont{{Jimenez}}},
  \bibinfo{journal}{\prd} \textbf{\bibinfo{volume}{66}},
  \bibinfo{pages}{063007} (\bibinfo{year}{2002}),
  \eprint{arXiv:astro-ph/0206014}.

\bibitem[{\citenamefont{{Fixsen} and {Mather}}(2002)}]{Fixsen2002}
\bibinfo{author}{\bibfnamefont{D.~J.} \bibnamefont{{Fixsen}}} \bibnamefont{and}
  \bibinfo{author}{\bibfnamefont{J.~C.} \bibnamefont{{Mather}}},
  \bibinfo{journal}{\apj} \textbf{\bibinfo{volume}{581}}, \bibinfo{pages}{817}
  (\bibinfo{year}{2002}).

\bibitem[{\citenamefont{{Olive} and {Skillman}}(2004)}]{Olive2004}
\bibinfo{author}{\bibfnamefont{K.~A.} \bibnamefont{{Olive}}} \bibnamefont{and}
  \bibinfo{author}{\bibfnamefont{E.~D.} \bibnamefont{{Skillman}}},
  \bibinfo{journal}{\apj} \textbf{\bibinfo{volume}{617}}, \bibinfo{pages}{29}
  (\bibinfo{year}{2004}), \eprint{arXiv:astro-ph/0405588}.

\end{thebibliography}

\end{document}